\documentclass[twocolumn]{aastex62}
\pdfoutput=1 %for arXiv submission
\usepackage{amstext,amsmath}
\usepackage[T1]{fontenc}
\usepackage[figure,figure*]{hypcap}
\usepackage{tabularx}
\usepackage{xcolor}

\usepackage{amsmath}

\renewcommand{\tabcolsep}{8mm}

\newcommand{\mL}{{\mathcal{L}}}
\newcommand{\ML}{{\text{ML}}}
\newcommand{\mM}{{\mathcal{M}}}
\newcommand{\bmM}{{\boldsymbol{\mathcal{M}}}}
\newcommand{\ba}{{\boldsymbol{a}}}
\newcommand{\bb}{{\boldsymbol{b}}}
\newcommand{\bd}{{\boldsymbol{d}}}
\newcommand{\bn}{{\boldsymbol{n}}}

\newcommand{\br}{{\boldsymbol{r}}}
\newcommand{\bs}{{\boldsymbol{s}}}
\newcommand{\bt}{{\boldsymbol{t}}}

\newcommand{\bw}{{\boldsymbol{w}}}
\newcommand{\bx}{{\boldsymbol{x}}}
\newcommand{\by}{{\boldsymbol{y}}}
\newcommand{\bz}{{\boldsymbol{z}}}
\newcommand{\bA}{{\boldsymbol{A}}}
\newcommand{\bB}{{\boldsymbol{B}}}
\newcommand{\bC}{{\boldsymbol{C}}}
\newcommand{\bD}{{\boldsymbol{D}}}
\newcommand{\bF}{{\boldsymbol{F}}}
\newcommand{\bG}{{\boldsymbol{G}}}
\newcommand{\bH}{{\boldsymbol{H}}}
\newcommand{\bI}{{\boldsymbol{I}}}

\newcommand{\bN}{{\boldsymbol{N}}}
\newcommand{\bS}{{\boldsymbol{S}}}
\newcommand{\bT}{{\boldsymbol{T}}}
\newcommand{\bU}{{\boldsymbol{U}}}
\newcommand{\bV}{{\boldsymbol{V}}}

\newcommand{\fg}{{\text{fg}}}
\newcommand{\fgo}{{\text{fg-only}}}
\newcommand{\balpha}{{\boldsymbol{\alpha}}}

\newcommand{\bdelta}{{\boldsymbol{\delta}}}
\newcommand{\bgamma}{{\boldsymbol{\gamma}}}
\newcommand{\bmu}{{\boldsymbol{\mu}}}
\newcommand{\bxi}{{\boldsymbol{\xi}}}
\newcommand{\bzeta}{{\boldsymbol{\zeta}}}
\newcommand{\bGamma}{{\boldsymbol{\Gamma}}}
\newcommand{\bDelta}{{\boldsymbol{\Delta}}}
\newcommand{\bLambda}{{\boldsymbol{\Lambda}}}
\newcommand{\bPsi}{{\boldsymbol{\Psi}}}
\newcommand{\bPhi}{{\boldsymbol{\Phi}}}
\newcommand{\bSigma}{{\boldsymbol{\Sigma}}}
\newcommand{\bzero}{{\boldsymbol{0}}}
\newcommand{\E}{{\text{E}}}
\newcommand{\Var}{{\text{Var}}}
\newcommand{\Cov}{{\text{Cov}}}
\newcommand{\Tr}{{\text{Tr}}}

\shorttitle{Examining the assumptions of global 21-cm signal analyses}
\shortauthors{Tauscher et al.}

\begin{document}

%\title{Critically examining the assumptions of global 21-cm signal analyses:
%
%False troughs can appear in single spectrum fits}

\title{Formulating and critically examining the assumptions of global 21-cm signal analyses:

How to avoid the false troughs that can appear in single spectrum fits}

\correspondingauthor{Keith Tauscher}

\author{Keith Tauscher}
\affiliation{Center for Astrophysics and Space Astronomy, Department of Astrophysical and Planetary Science, University of Colorado, Boulder, CO 80309, USA}
\affiliation{Department of Physics, University of Colorado, Boulder, CO 80309, USA}

\author{David Rapetti}
\affiliation{NASA Ames Research Center, Moffett Field, CA 94035, USA}
\affiliation{Universities Space Research Association, Mountain View, CA 94043, USA}
\affiliation{Center for Astrophysics and Space Astronomy, Department of Astrophysical and Planetary Science, University of Colorado, Boulder, CO 80309, USA}

\author{Jack~O.~Burns}
\affiliation{Center for Astrophysics and Space Astronomy, Department of Astrophysical and Planetary Science, University of Colorado, Boulder, CO 80309, USA}

\email{Keith.Tauscher@colorado.edu}

\begin{abstract}
The assumptions inherent to global 21-cm signal analyses are rarely delineated.~In this paper, we formulate a general list of suppositions underlying a given claimed detection of the global 21-cm signal.~Then, we specify the form of these assumptions for two different analyses: 1) the one performed by the EDGES team showing an absorption trough in brightness temperature that they modeled separately from the sky foreground and 2) a new, so-called Minimum Assumption Analysis (MAA), that makes the most conservative assumptions possible for the signal. We show fits using the EDGES analysis on various beam-weighted foreground simulations from the EDGES latitude with no signal added. Depending on the beam used, these simulations produce large false troughs due to the invalidity of the foreground model to describe the combination of beam chromaticity and the shape of the Galactic plane in the sky, the residuals of which are captured by the ad hoc flattened Gaussian signal model. On the other hand, the MAA provides robust fits by including many spectra at different time bins and allowing any possible 21-cm spectrum to be modeled exactly.~We present uncertainty levels and example signal reconstructions found with the MAA for different numbers of time bins.~With enough time bins, one can determine the true 21-cm signal with the MAA to $<10$ times the noise level.
\end{abstract}


\section{Introduction}

The highly redshifted 21-cm signal generated by the hyperfine, spin-flip transition of neutral hydrogen tracks the history of the Universe after recombination and before reionization \citep{Pritchard:12}, from the Dark Ages, before there were any compact sources, through Cosmic Dawn, when the first stars were born. This signal comes in two forms, the power spectrum, which has both angular and spectral dependence, and the sky-averaged global signal, which is a single spectrum. While the former is being focused on by experiments like the Hydrogen Epoch of Reionization Array \citep[HERA,][]{HERA:16}, the former, which is the focus of this paper, is actively being searched for by experiments such as the Large-Aperture Experiment to Detect the Dark Age \citep[LEDA,][]{Bernardi:18}, the Shaped
Antenna measurement of the background RAdio Spectrum \citep[SARAS,][]{Singh:18}, the Cosmic Twilight Polarimeter~\citep[CTP,][]{Nhan:19}, the Radio Experiment for the Analysis of Cosmic Hydrogen~\citep[REACH,][]{Lera:19}, and the Experiment to Detect the Global Epoch-of-Reionization Signature \citep[EDGES,][]{Monsalve:17b,Monsalve:18}.~In particular, in \cite{Bowman:18} the EDGES team reported a trough in the sky-averaged radio spectrum at 78 MHz using an analysis technique that we will examine in detail in this paper. Space-based mission concepts, such as the Dark Ages Polarimeter PathfindER \citep[DAPPER;][]{Burns:19,Burns:20} and its precursor the Dark Ages Radio Explorer \citep[DARE,][]{Burns:17}, are also being developed to measure the global 21-cm signal from the vicinity of the Moon, where experiments do not have to deal with systematic effects such as human-generated Radio Frequency Interference (RFI) or attenuation and emission from the ionosphere or interactions of the antenna beam with the environment.

While all of these experiments must be designed carefully to have the sensitivity to measure the global signal, which is expected to be a few hundred mK deep, this paper will focus on the data analysis techniques necessary to extract the signal from the beam-weighted galactic foreground emission, which, at the relevant frequencies, is generally at a level of a few thousand K \citep[for the magnitude of the beam-weighted foreground seen by EDGES, see Figure 1a of][]{Bowman:18}. The shape of the beam-weighted foreground in the absence of the global signal is not known a priori, so simple subtraction is not an option. Therefore, one must devise a model which is known (or at least reasonably assumed) to encapsulate the possible values of the beam-weighted foreground.~This task would be easy if the beam would be achromatic, but such beams do not exist. Real antennas have beams that change with frequency over the wide bandwidth that these experiments need to measure.~This beam chromaticity distorts the spectral shapes of the intrinsic foreground as measured by an achromatic beam (and which would be well fit by models such as the ones EDGES uses) in ways particular to the antenna being used. Therefore, to fit the beam-weighted foreground, a model that is particular to the given antenna and experiment location must be created. Such a model can be created via Singular Value Decomposition (SVD) of a training set as it is in the pipeline laid out in \cite{Tauscher:18}, \cite{Rapetti:19}, and \cite{Tauscher:20}. The method presented for making this model is similar in spirit to that defined in \cite{Switzer:14}, except the foreground basis is derived from an a priori training set instead of the data itself. It is also very similar to the method performed by \cite{Vedantham:14}, although in that paper, SVD is performed on spectra at different time bins to define modes that span frequencies, while in this paper, different simulations of spectra defined at multiple time bins are used to define modes that span both frequencies and time bins. This difference is crucial because the constraining power of our method comes from the fact that its foreground model encodes correlations between time bins, which is not true of the purely spectral modes arising from the methods of \cite{Vedantham:14}. Our method, a new variant of which will be described in Section~\ref{sec:minimum-assumption-analysis}, connects with the idea from \cite{Liu:13} that angular information can effectively supplement spectral information in signal estimation. However, instead of requiring instruments with angular resolution, our method gleans spatial information from time-dependent drift-scan measurements from single antenna instruments with no angular resolution, where the spectra weight the entire spatial sky into a single pixel.

This paper is especially relevant in light of the large volume of literature written by the community in their attempts to explain the trough presented by \cite{Bowman:18}, especially its depth. Some have explored the possibility that it could represent a larger radio background than expected \citep{Feng:18,Ewall-Wice:18,Ewall-Wice:19,Fialkov:19,Mebane:19}, while others have hypothesized that Rutherford-like scattering of dark matter could cause the hydrogen gas to cool faster than adiabatic cooling would imply \citep{Munoz:18,Barkana:18,Barkana:18b,Fialkov:18,Loeb:18,Berlin:18}. While none of these ideas perfectly describe the detection \citep[see, e.g.][for a criticism of dark matter explanations]{CS:19}, they illustrate the temptation to use the results of \cite{Bowman:18} to explore possible exotic physics. In addition to these physical explanations, some have explored possible unmodeled systematic effects present in the data \citep{Hills:18,Bradley:2019,Draine:18,Singh:19,Spinelli:19,Sims:20}. In this paper, we layout yet another possible explanation of the EDGES results---that the modeling performed could be mistaking beam cromaticity distortion of the foreground for signal.

In Section~\ref{sec:enumerating-assumptions}, we lay out the general presumptions necessary to perform any global 21-cm signal experiment. In Section~\ref{sec:edges-mock-simulations-and-analysis}, we generate and fit multiple EDGES-like simulations of beam-weighted foregrounds with the analysis method of \cite{Bowman:18}. In Section~\ref{sec:minimum-assumption-analysis}, we propose a new, minimum assumption analysis that allows for any possible 21-cm global signal and demonstrate it on simulated spectra with a signal and beam-weighted foreground. In Section~\ref{sec:discussion}, we discuss and compare the results of the two analyses. Finally, we conclude in Section~\ref{sec:conclusions}.

\section{Enumerating assumptions} \label{sec:enumerating-assumptions}

The likelihood used when doing a foreground-only fit \citep[like the one done to generate the residuals found in Figure 1b of][]{Bowman:18} is
\begin{equation}
  \mL_\fgo(\ba) \propto e^{-(\by-\bF\ba)^T\bC^{-1}(\by-\bF\ba)/2},
\end{equation}
where $\by$ is the spectrum written as a column vector, $\bC$ the covariance matrix of the data's noise distribution, and $\bF$ a matrix with the basis vectors used to fit the foreground as columns. The value of $\ba$ that maximizes $\mL_\fgo(\ba)$ is
\begin{equation}
  \ba_{\ML,\fgo} = (\bF^T\bC^{-1}\bF)^{-1}\bF^T\bC^{-1}\by \label{eq:fg-only-maximum-likelihood-coefficients}
\end{equation}
and the maximum likelihood reconstruction of the foreground, $\bgamma_\fg$, is given by $\bgamma_\fg=\bF\ba_{\ML,\fgo}$, or $\bgamma_\fg=\bPhi\by$, where
\begin{equation}
  \bPhi = \bF(\bF^T\bC^{-1}\bF)^{-1}\bF^T\bC^{-1}
\end{equation}
is the matrix that projects vectors into the column space of $\bF$ (i.e.~the space of spectra formable by the linear foreground model). The residual, $\br$, unfit by this procedure is given by $\br=\by-\bgamma_\fg = (\bI-\bPhi)\by$. The data are given by a sum of a foreground term, $\by^{(\fg)}$, a signal term $\by^{(21)}$, and random noise $\bn$.

We can write $\by^{(\fg)}=\bF\balpha+\bdelta$ where $\balpha$ is the coefficient vector that best describes $\by^{(\fg)}$ given the basis matrix $\bF$\footnote{If the signal, $\by^{(21)}$, was zero and there was no random noise, then $\balpha$ would be equal to $\ba_{\ML,\fgo}$. However, in the general case where there is a nonzero signal, $\ba_{\ML,\fgo}$ absorbs some of the signal spectrum. In addition, $\ba_{\ML,\fgo}$ is subject to small biases caused by random noise, whereas in this formulation $\balpha$ is the best fitting coefficient vector of the ideal, noiseless foreground (i.e. the one which would be measured given infinite integration time with a stable instrument).} and $\bdelta$ is the part of $\by^{(\fg)}$ that cannot be fit by the foreground model, i.e. the part of $\by^{(\fg)}$ that is not in the column space of $\bF$, which satisfies $\bPhi\bdelta=\bzero$. This means that $(\bI-\bPhi)\by^{(\fg)}=\bdelta$ and the foreground-only residual $\br$ is given by
\begin{equation}
    \br = (\bI-\bPhi)(\by^{(21)}+\bdelta+\bn). \label{eq:fg-only-residual}
\end{equation}
Therefore, to solve for the signal plus foreground inaccuracy and noise, which we define as $\bs=\by^{(21)}+\bdelta+\bn$, we must find the general solution of $(\bI-\bPhi)\bs=\br$, which is the sum of any particular solution, $\bs_p$, and the general solution, $\bs_h$, to the homogeneous equation, $(\bI-\bPhi)\bs_h=\bzero$. We guess that $\br$ is a particular solution and verify by noting that $(\bI-\bPhi)\br=(\bI-\bPhi)^2\by=(\bI-2\bPhi+\bPhi^2)\by=(\bI-\bPhi)\by=\br$ because $\bPhi^2=\bPhi$. The homogeneous solution is any vector in the null space of $\bI-\bPhi$, which is equivalent to any vector that remains unchanged when multiplied on the left by $\bPhi$. Since $\bPhi$ is a projection matrix onto the columns of $\bF$, any vector in the column space of $\bF$ remains unchanged when being multiplied by $\bPhi$. Therefore, $\bs_h=\bF\bx$ for any $\bx\in\mathbb{R}^N$. This means that the signal satisfies
\begin{equation}
  \by^{(21)} = \br+\bF\bx-\bdelta-\bn \label{eq:non-unique-solution}
\end{equation}
for some unknown vector $\bx$. Equation~\ref{eq:non-unique-solution} implies that, due to the filtering out of the foregrounds when fitting, the signal can have any foreground modes added to it without changing the quality of the fit. \textbf{This shows that with only one spectrum and no more information, even if $\bdelta=\bzero$ (i.e. the foreground model is perfect), the 21-cm signal cannot be uniquely determined.} Thus, an extra assumption about the form of the signal is required. The full set of assumptions needed to extract the global 21-cm signal with useful, rigorous uncertainties is as follows:
\begin{enumerate}
  \item Sky-averaged radio data contains a sum of beam-weighted foreground emission and the global signal.~This assumes a well-calibrated instrument. \label{item:calibration-assumption}
  \item The noise of the data follows a known or estimated distribution.\footnote{Usually, this distribution is a zero-mean Gaussian specified entirely by its covariance matrix, $\bC$, which must be known up to a constant of proportionality. The proportionality constant must be sufficiently close to 1 if we also wish to measure goodness-of-fit.} \label{item:noise-assumption}
  \item The true beam-weighted foreground can be fit with the given foreground model, described by the basis matrix $\bF$, to well below the noise level of the data. This is equivalent to $\bdelta^T\bC^{-1}\bdelta\ll N_c$, where $N_c$ is the number of channels in $\by$ and $\bdelta$ is the unmodeled component of the foreground as above, given by $\bdelta=(\bI-\bPhi)\by^{(\fg)}$. \label{item:foreground-assumption}
  \item The signal follows a specific form. \label{item:signal-assumption}
\end{enumerate}
Assumption~\ref{item:signal-assumption} can take many forms, some much stronger and more unjustified than others. In \cite{Bowman:18}, the signal is assumed to follow the form of a flattened Gaussian profile. Note that this is an assumption for analyzing the data and cannot be justified by examining the data itself, especially in light of the Bayesian evidence-based analysis of \cite{Sims:20}, which showed that many possible assumed signal models lead to essentially the same Bayesian evidence. Crucially, the errors obtained via fitting a given model of the signal do not account for the unknown likelihood that the true signal can be fit by that model.

The specific flattened Gaussian profile in \cite{Bowman:18} merely represents the value of $\by^{(21)}$ from Equation~\ref{eq:non-unique-solution} that best matches the assumed flattened Gaussian model. Different values of $\bx$ along with different assumed signal models lead to equally valid interpretations of the data \citep[see, e.g.][]{Hills:18,Bradley:2019,Singh:19,Sims:20}.

\section{EDGES-like analysis} \label{sec:edges-mock-simulations-and-analysis}

\subsection{Assumptions}

\subsubsection{Assumption~\ref{item:calibration-assumption}: calibration}

The EDGES team has worked extensively on the calibration of their receiver \citep{Monsalve:17b,Monsalve:17}. Assuming solar, weather, RFI, and other transient events are removed, due to the rigor of their lab measurements and calibration strategy, it is reasonable to assume that, to the mK level, their data consists solely of beam-weighted foreground and global 21-cm signal.

\subsubsection{Assumption~\ref{item:noise-assumption}: noise level} \label{sec:edges-noise-assumption}

Although the covariance distribution of the noise is not known, the residuals presented in \cite{Bowman:18} seem to have a relatively flat magnitude of 20 mK when many smooth modes are removed. So, for our simulations we use $\bC=\sigma^2\bI$ where $\sigma=20$ mK and $\bI$ is the identity matrix.

\subsubsection{Assumption~\ref{item:foreground-assumption}: foreground model}

The EDGES analysis involves multiple linear foreground models, but a common model in the literature is a polynomial in $\ln{(\nu/\nu_0)}$ multiplied by $(\nu/\nu_0)^{-2.5}$ where $\nu$ is the observed frequency and $\nu_0$ is a reference frequency,
\begin{equation}
  \mM_\fg(a_1,a_2,\ldots,a_{N_t}) = \left(\frac{\nu}{\nu_0}\right)^{-2.5} \sum_{k=1}^{N_t}a_k\left[\ln{\left(\frac{\nu}{\nu_0}\right)}\right]^{k-1}. \label{eq:edges-foreground-model}
\end{equation}
This is equivalent to assuming that $\bF$ is a matrix with the terms $(\nu/\nu_0)^{-2.5}[\ln{(\nu/\nu_0)}]^{k-1}$ as its columns and the foreground coefficient vector $\ba$ is given by $\ba^T=\begin{bmatrix} a_1 & a_2 & \cdots & a_{N_t} \end{bmatrix}$. This model is based on the Taylor series approximation of $T_0(\nu/\nu_0)^\beta$ around $\beta=-2.5$, which should adequately describe the intrinsic synchrotron foreground in each sky direction (see also Hibbard et al., in preparation).

\subsubsection{Assumption~\ref{item:signal-assumption}: signal model} \label{sec:edges-signal-assumption}

In \cite{Bowman:18}, the EDGES team uses a flattened Gaussian signal model of the form
\begin{equation}
  \bmM_{21}(A,\mu,w,\tau) = A\left(\frac{1-e^{-\tau e^D}}{1-e^{-\tau}}\right),
\end{equation}
where
\begin{equation}
    D = \left[\frac{2(\nu-\mu)}{w}\right]^2\ \ln{\left\{-\frac{1}{\tau}\ln{\left[\frac{1+e^{-\tau}}{2}\right]}\right\}}.
\end{equation}
In this section, we adopt the same model.

\begin{figure*}
  \centering
  \includegraphics[width=0.48\textwidth]{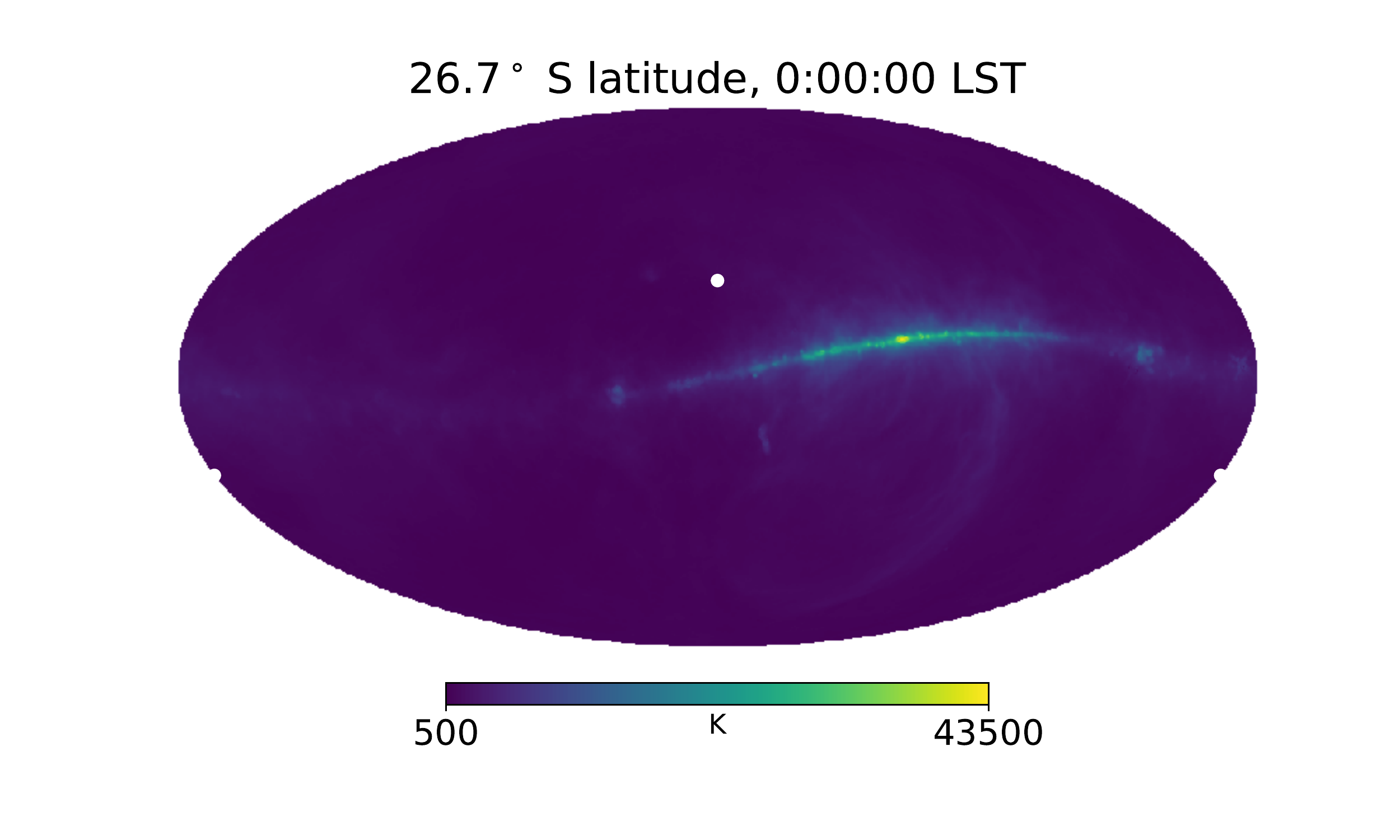}
  \includegraphics[width=0.48\textwidth]{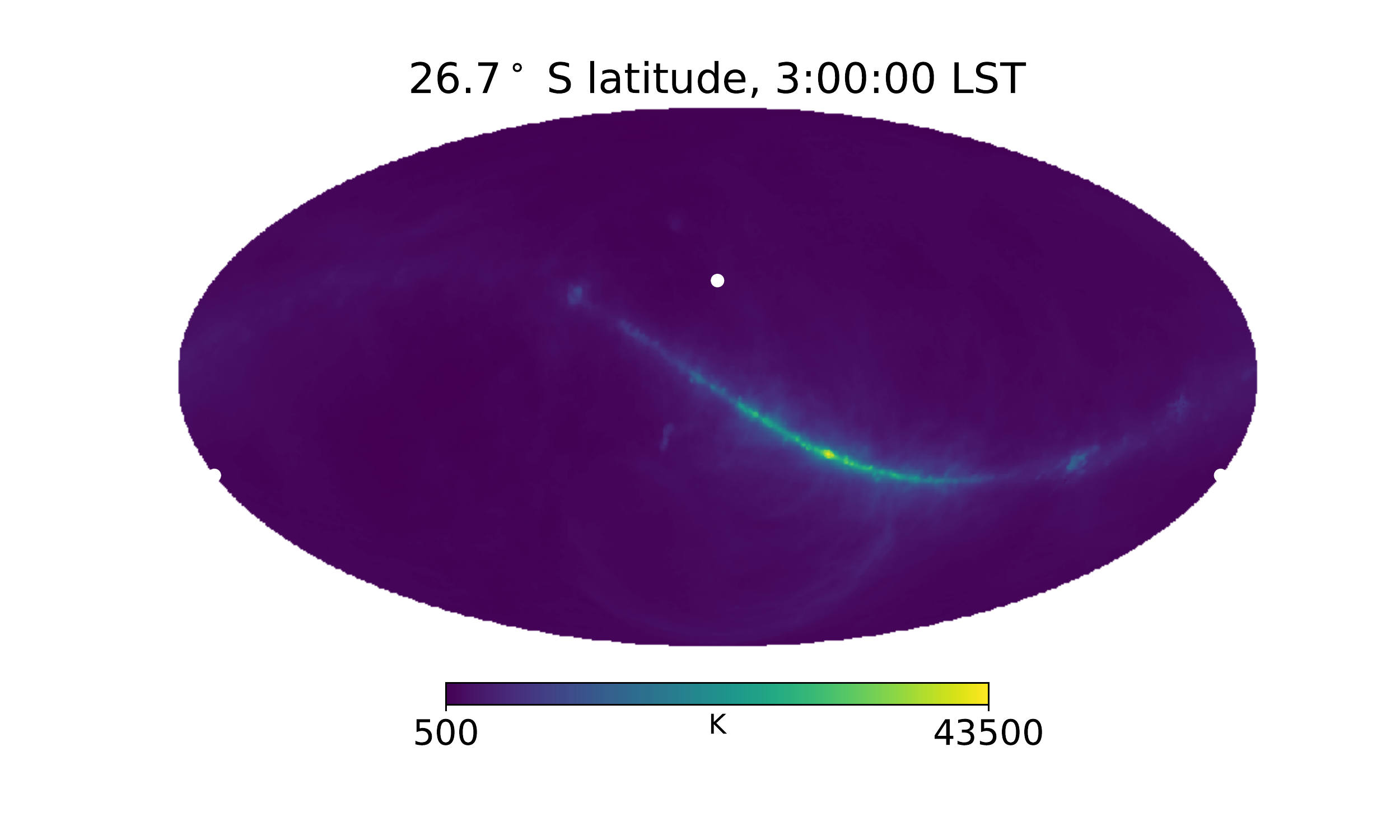}
  \includegraphics[width=0.48\textwidth]{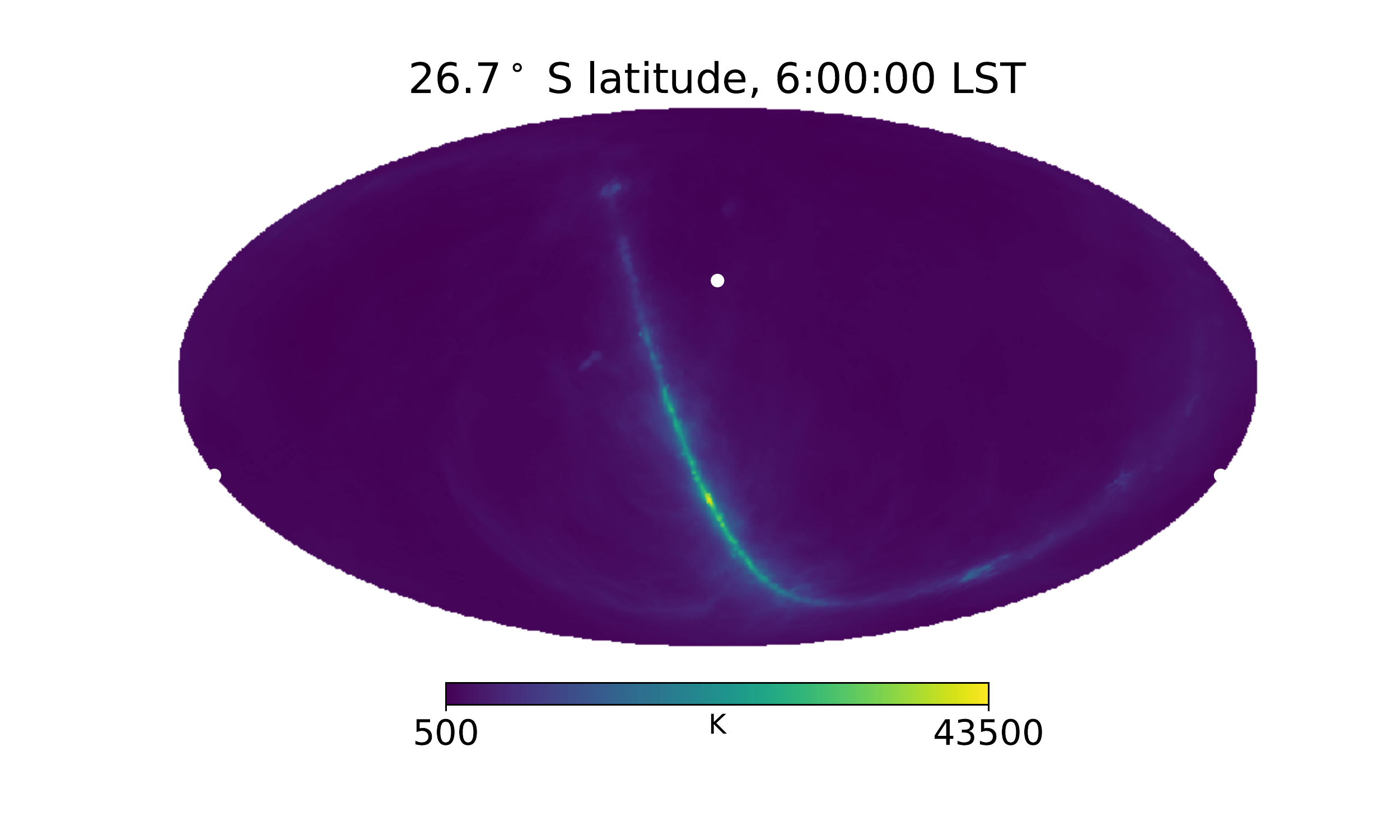}
  \includegraphics[width=0.48\textwidth]{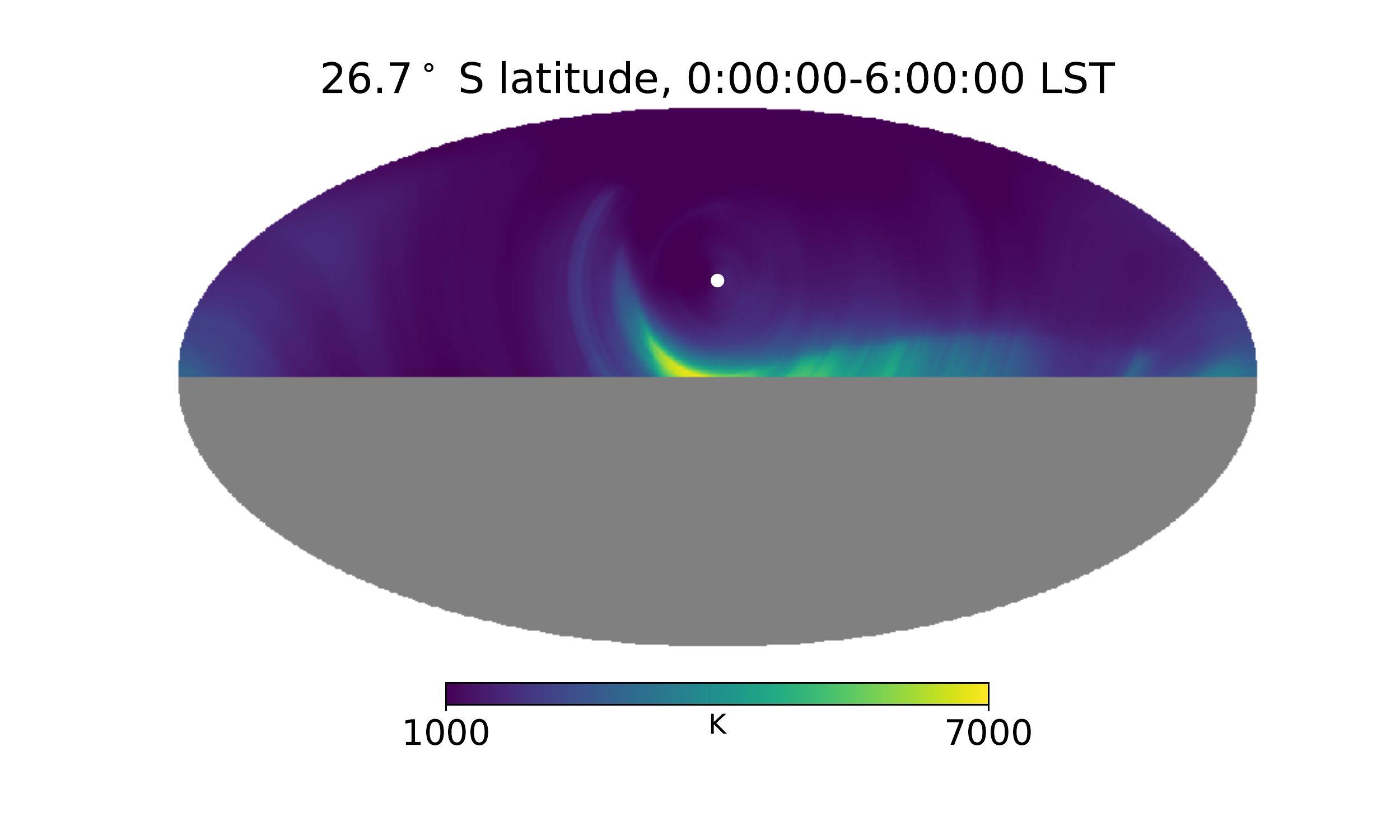}
  \caption{The Haslam map scaled to 80 MHz with a spectral index of -2.5 shown with the zenith from the EDGES latitude ($26.7^\circ$ S) as the uppermost point at different LSTs. The top left, top right, and lower left panels show the cases at LST hours of 0, 3, and 6. The lower right panel shows the case where the map has been smoothly smeared between hours 0 and 6 with directions below the horizon (grey region) masked out, which is the foreground used in the simulations in Sections~\ref{sec:edges-mock-simulations-and-analysis}~and~\ref{sec:minimum-assumption-analysis}. In all panels, the celestial poles are marked by white dots.} \label{fig:edges-simulation-foreground}
\end{figure*}

\subsection{Simulations} \label{sec:edges-simulations}

To illustrate the potential problems with the EDGES analysis in \cite{Bowman:18}, we have created simple simulations of the beam-weighted foreground expected to be seen from the Murchison Radio-astronomy Observatory (MRO) where the experiment is located.\footnote{$26.7^\circ$ S, $116.6^\circ$ E} This section will lay out those simulations as well as apply the EDGES analysis to them.

The foreground brightness temperature, $y^{(\fg)}$, in the simulations presented throughout Sections~\ref{sec:edges-mock-simulations-and-analysis}~and~\ref{sec:minimum-assumption-analysis} is given by
\begin{equation}
  y^{(\fg)}(\nu) = \int B(\nu,\theta,\phi)\ T(\nu,\theta,\phi)\ d\Omega,
\end{equation}
where $B(\nu,\theta,\phi)$ is the antenna beam, $T(\nu,\theta,\phi)$ is the foreground emission, and $\theta$ and $\phi$ are the spherical coordinate angles.

\begin{figure*}[t!]
  \centering
  \includegraphics[width=0.98\textwidth]{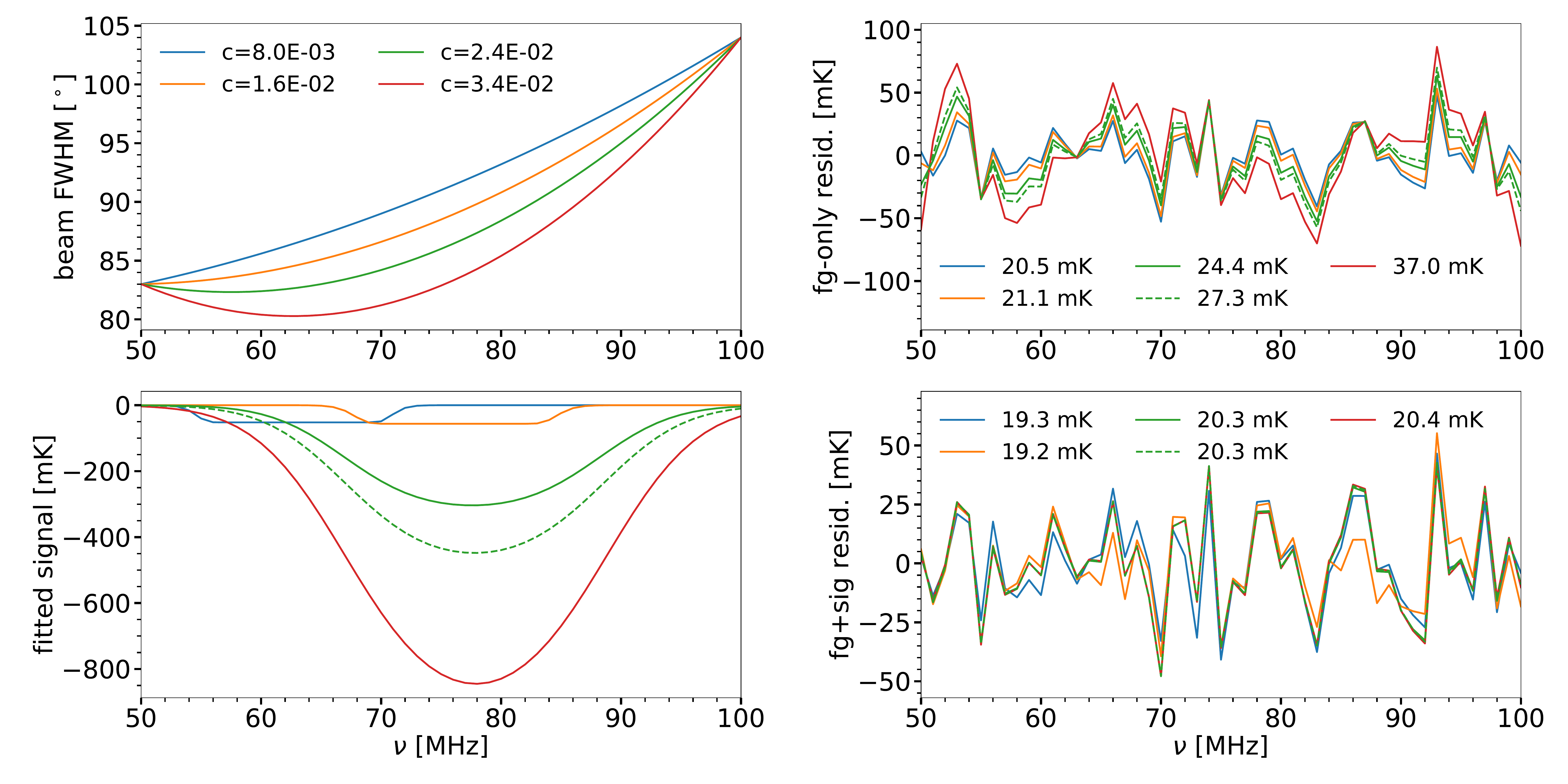}
  \caption{Summary of EDGES-like simulations and analyses.~Throughout the figure, curves with the same color correspond to the same beam FWHM curvature. \textit{Upper left}: FWHM curves of the four simulated antenna beams as a function of frequency. \textit{Upper right}: Residuals when the foregrounds generated with the beams from LST hours 0-6 at the EDGES site are fit with the foreground model of Equation~\ref{eq:edges-foreground-model} with $N_t=6$. The Root-Mean-Square (RMS) value of each residual is shown in the legend (to be compared to noise level of 20 mK). \textit{Lower left}: Flattened Gaussian best fits for each beam case when the foreground and absorption trough models are fit simultaneously. Large troughs appear as the beam curvature grows. \textit{Lower right}: Residuals of the combined foreground and absorption trough fits that produced the signals in the lower left panel. The green dashed lines in the upper right and lower panels show results found when the Haslam map is replaced by a toy model of the galaxy that accounts only for the shape of the galactic plane (see Section~\ref{sec:toy-galaxy-model} for details on the model). Note that the same noise realization was added to all five simulated spectra for comparison purposes.}
  \label{fig:edges-simulations}
\end{figure*}

\subsubsection{Antenna beam}

The antenna beam used in our simulations is a Gaussian beam with a frequency dependent Full Width at Half Maximum (FWHM), $\alpha(\nu)$, i.e.
\begin{equation}
  B(\nu,\theta,\phi) \propto \exp{\left\{-4\ln{2}\left[\frac{\theta}{\alpha(\nu)}\right]^2\right\}}.
\end{equation}
The FWHMs we use are parabolas with fixed endpoints at $(50\text{ MHz}, 83^\circ)$ and $(100\text{ MHz}, 104^\circ)$\footnote{These endpoints are similar to those of the beam of the blade antenna used by EDGES \citep[see extended data figure 4a of][]{Bowman:18}.} and varying curvature, $c$. This means, they have the form
\begin{equation}
  \alpha(\nu) = \alpha_l(\nu) + c\alpha_c(\nu),
\end{equation}
where $\alpha_l(\nu)$ is the line connecting the two endpoints and $\alpha_c(\nu)$ is the parabola with unit curvature and zeros at the endpoint frequencies. $\alpha_l(\nu)$ and $\alpha_c(\nu)$ are given by
\begin{subequations}
\begin{align}
  \alpha_l(\nu) &= 0.42^\circ\left(\frac{\nu}{1\text{ MHz}}\right)+62^\circ ,\\
  \alpha_c(\nu) &= \frac{1}{2}\nu^2 - (75\text{ MHz})\nu+(50\text{ MHz})^2 .
\end{align}
\end{subequations}
For illustration purposes, we use $c/(1^\circ/\text{MHz}^2)$ values of $8.0\times 10^{-3}$, $1.6\times 10^{-2}$, $2.4\times 10^{-2}$, and $3.4\times 10^{-2}$. The FWHM curves of the four beams used in our simulations are shown in the upper left panel of Figure~\ref{fig:edges-simulations}. To ensure that results are not tainted by any fraction of the beam below the horizon, the beams are normalized such that
\begin{equation}
  \iint_{\mathcal{H}} B(\nu,\theta,\phi)\ d\Omega = 1,
\end{equation}
where $\mathcal{H}$ represents the half of the sphere above the horizon, i.e. $0\le\theta\le\pi/2$ and $0\le\phi<2\pi$.

\subsubsection{Foreground map}

For the simulations used in this paper, we use the 408 MHz map from \cite{Haslam:82}, scaled to lower frequencies via multiplication by a power law with spectral index -2.5, i.e.
\begin{equation}
    T(\nu,\theta,\phi,t) = T_{\text{Haslam}}(\theta,\phi,t)\times\left(\frac{\nu}{408\text{ MHz}}\right)^{-2.5}.
\end{equation}
The time dependence indicated in both $T$ and $T_\text{Haslam}$ is determined by the rotation of the Earth from the EDGES latitude as a function of sidereal time.~The Haslam maps, rotated to match the zenith direction at 0, 3, and 6 hr LST (local sidereal time) at $26.7^\circ$ S latitude, are shown in the top left, top right, and bottom left panels of Figure~\ref{fig:edges-simulation-foreground}.

\subsubsection{Observation strategy}

The simulations were performed by averaging together equally all beam-weighted foregrounds between LST hours 0-6 from the EDGES latitude.~As the central bulge of the Milky Way is closest to the zenith between LST hours 17 and 18, this strategy amounts to averaging during low foreground times. This observation strategy leads to an effective foreground given by
\begin{equation}
    T_{\text{eff}}(\nu,\theta,\phi,0\rightarrow 6\text{ hr LST}) = \int_{0\text{ hr LST}}^{6\text{ hr LST}} T(\nu,\theta,\phi,t)\ dt
\end{equation}
and shown at 80 MHz in the bottom right panel of Figure~\ref{fig:edges-simulation-foreground}.

\subsubsection{No 21-cm signal added}

No 21-cm global signal was included in the simulations.~All fits shown in Figure~\ref{fig:edges-simulations} are performed on the beam-weighted foreground alone with noise.

\subsubsection{Random noise}

To each simulated measurement of the beam-weighted foreground we add the same realization of noise at the 20 mK level (see Section~\ref{sec:edges-noise-assumption}) to control for the effect of noise without ignoring it.

\subsection{Likelihood maximization techniques}

For foreground-only fits, we maximize the likelihood through the analytical computation of the coefficient vector $\ba_{\ML,\fgo}$ as in Equation~\ref{eq:fg-only-maximum-likelihood-coefficients}. Then, the maximum likelihood foreground reconstruction is simply $\bF\ba_{\ML,\fgo}$.

For fits with both the linear foreground model and the flattened Gaussian absorption trough model, we numerically explore the space of the trough parameters $\{A,\mu,w,\tau\}$ to find the maximum likelihood value.\footnote{We use the \texttt{minimize} function from \texttt{scipy.optimize} to minimize the negative log-likelihood.} At the $k^{\text{th}}$ iteration of the optimization algorithm, the exploration is at $(A^{(k)},\mu^{(k)},w^{(k)},\tau^{(k)})$. To evaluate the full likelihood at that point, we perform a foreground-only fit, as above, on the modified data spectrum given by $\by-\bmM_{21}(A^{(k)},\nu^{(k)},w^{(k)},\tau^{(k)})$. This allows for exploration of only the nonlinear absorption trough parameters instead of including the foreground parameters, which would slow down the analysis. This technique is similar to that used in \cite{Rapetti:19}, except instead of exploring the entire posterior distribution of the trough parameters with a Markov Chain Monte Carlo simulation, here we merely wish to find its maximum through gradient ascent.

\subsection{EDGES-like analysis results}

The beam-weighted foregrounds with noise from the simulations described in Section~\ref{sec:edges-simulations} were fit with the model given by Equation~\ref{eq:edges-foreground-model}, which is meant to describe the sky foreground completely.~The residuals from these foreground-only fits are shown in the upper right panel of Figure~\ref{fig:edges-simulations}. While the beam-weighted foreground from the lowest curvature FWHM beam case is fit to the noise level, it is clear that the residuals grow as the curvature increases, with the largest curvature beam case yielding residuals two times higher than the noise level. These foreground-only residuals correspond to the $\br$ curves from Equation~\ref{eq:fg-only-residual} with $\by^{(21)}=\bzero$.

As in \cite{Bowman:18}, we also performed fits that include both the linear foreground model and a flattened Gaussian absorption trough 
model. The resulting troughs are shown in the lower left panel of Figure~\ref{fig:edges-simulations}. As the beam FWHM curvature increases, the size of the fit trough increases up to $\sim$800 mK, showing that inaccuracies in the foreground model (i.e.~violations of assumption~\ref{item:foreground-assumption}) can lead to false troughs appearing in fits. The residuals to these foreground and trough fits are shown in the lower right panel of Figure~\ref{fig:edges-simulations}. In all four cases, these residuals are at the 20 mK noise level, showing that flattened Gaussian troughs can effectively complement the smooth foreground model from Equation~\ref{eq:edges-foreground-model} to fit the foreground down to the noise level.

The simulations presented here show that some frequency dependencies, such as curvature in the FWHM of an antenna beam \citep[which there is some sign of in extended data Figure 4a of][although only three frequencies are shown]{Bowman:18}, can induce structure in the beam-weighted foreground that cannot be fit out by a $6^{\text{th}}$ order polynomial. Not accounting for this structure can lead to artificial troughs being found when a signal model is fit simultaneously with the beam-weighted foreground. It is important to note that the simulations proposed here use simple beams, fully characterized by the FWHM function $\alpha(\nu)$, whereas real antenna beams have many independent modes of variation corresponding to the geometry and electrical properties of antenna components. One complication in the specific case of the EDGES beam is the lack of azimuthal symmetry, leading to different E- and H-plane beam patterns.

\section{Minimum assumption analysis} \label{sec:minimum-assumption-analysis}

\subsection{A more robust assumption~\ref{item:signal-assumption}} \label{sec:minimum-signal-assumption}

For a robust analysis, instead of assumption~\ref{item:signal-assumption} as laid out for the EDGES-like analysis in Section~\ref{sec:edges-signal-assumption}, which is not properly motivated, we can utilize exclusively the fact---not even used in the previous analysis---that, by definition, the global 21-cm signal must be spatially uniform. If the data $\by$ is a concatenation of $N_s$ spectra taken by the same instrument instead of a single spectrum, i.e.
\begin{equation}
    \by = \begin{bmatrix} \by_1^T & \by_2^T & \cdots & \by_{N_s}^T \end{bmatrix}^T,
\end{equation}
then the spatial uniformity of the signal implies that the signal contribution to each spectrum is identical, i.e. $\by^{(21)}_k=\by^{(21)}$ for all $k$. On the other hand, the foregrounds of each spectrum will be different (unless the same sky is overhead in two or more of the spectra).

The analysis with the fewest possible assumptions about the signal would involve assuming nothing about the spectral behavior of the signal, i.e.~using a signal model that can fit all values of $\by^{(21)}$ in $\mathbb{R}^{N_\nu}$ where $N_\nu$ is the number of frequencies.~This can be achieved by computing a so-called ``expansion matrix'' $\bPsi$ \citep[see][for the initial definition]{Tauscher:18} that encodes how the signal appears in the data, i.e. the signal term in the full data, $\by$, is $\bPsi\by^{(21)}$. To encode the defining information aforementioned that all spectra have the same signal in them, we use $\bPsi=\begin{bmatrix} \bI & \bI & \cdots & \bI \end{bmatrix}^T$. We term an analysis with this form of the signal the minimum assumption analysis (MAA).

$\bPsi$ can also be adapted to fit different experimental designs, such as full Stokes measurements (e.g., from CTP and DAPPER) or data for which a different amount of sky is blocked in each spectrum (such as it can be for DAPPER due to the shifting position of the moon).

\subsection{Assumption~\ref{item:foreground-assumption}: choosing foreground basis}

\subsubsection{Polynomial bases}

Using identical and independent basis sets for each spectrum leads to a degeneracy between the foreground and signal models, causing infinite uncertainties. To see this, suppose that each spectrum has its own polynomial basis, represented by the matrix $\bF_0$. In this case, the full foreground basis matrix is
\begin{equation}
  \bF = \begin{bmatrix} \bF_0 & \bzero & \cdots & \bzero \\ \bzero & \bF_0 & \cdots & \bzero \\ \vdots & \vdots & \ddots & \vdots \\ \bzero & \bzero & \cdots & \bF_0 \end{bmatrix}.
\end{equation}
Multiplying this basis by a coefficient vector of the form $\bx^{(\fg)}=\begin{bmatrix} \bzeta^T & \bzeta^T & \cdots & \bzeta^T \end{bmatrix}^T$, leads to a foreground vector of the form $\bF\bx^{(\fg)}=\bPsi\bF_0\bzeta$. Since, in this case, the foreground basis can generate vectors in the column space of $\bPsi$, the foreground and signal bases are degenerate and the uncertainties diverge to infinity. This is true even if the $\bF_0$ basis can exactly fit the foregrounds in each spectrum. Note that this explanation holds no matter what the true beam-weighted foreground is because it only depends on the form of the beam-weighted foreground model.

\subsubsection{General choice of basis} \label{sec:choosing-foreground-basis}

We wish to find a basis that satisfies $\bF^T\bC^{-1}\bF=\bI$ and can accurately fit the beam-weighted foreground expected for each spectrum. We suggest generating a simulated training set of foregrounds of the form
\begin{equation}
  \bB = \begin{bmatrix} \bb^{(1)}_1 & \bb^{(2)}_1 & \cdots & \bb^{(N_t)}_1 \\   \bb^{(1)}_2 & \bb^{(2)}_2 & \cdots & \bb^{(N_t)}_2 \\ \vdots & \vdots & \ddots & \vdots \\ \bb^{(1)}_{N_s} & \bb^{(2)}_{N_s} & \cdots & \bb^{(N_t)}_{N_s} \end{bmatrix},
\end{equation}
where $\bb^{(m)}_n$ is the $n^{\text{th}}$ spectrum of the $m^{\text{th}}$ simulation and there are $N_t$ total simulations. For a given number of basis vectors $N_F$, weighted Singular Value Decomposition (SVD) of $\bB$ as in \cite{Tauscher:18} yields a basis $\bF$ satisfying the normalization conditions.\footnote{Weighted SVD refers to a decomposition of the form $\bB=\bU\bSigma\bV^T$ where $\bU^T\bC^{-1}\bU=\bI$, $\bV^T\bV=\bI$, and $\bSigma$ is a rectangular diagonal matrix with decreasing non-negative elements on the diagonal. The basis matrix $\bF$ is made from the first $N_F$ columns of $\bU$.} This basis will encode expected correlations between the different spectra in the foreground model directly, which is a crucial aspect of the analysis necessary to obtain finite errors \citep[see also][]{Tauscher:20} when allowing for any signal to be fit as discussed in Section~\ref{sec:minimum-signal-assumption}.

\subsection{Maximum likelihood calculations}

With a data vector $\by$, a signal expansion matrix $\bPsi$, a noise covariance matrix $\bC$, and a foreground basis matrix $\bF$, the channel covariance $\bDelta^{(21)}$ of the signal and the corresponding mean $\bgamma^{(21)}$ are given by
\begin{subequations}
\begin{align}
  \bDelta^{(21)} &= \left[\bPsi^T\bC^{-1}(\bI-\bPhi)\bPsi\right]^{-1}, \label{eq:signal-channel-covariance-in-terms-of-expansion-matrix} \\
  \bgamma^{(21)} &= \bDelta^{(21)}\bPsi^T\bC^{-1}(\bI-\bPhi)\by, \label{eq:signal-channel-mean-in-terms-of-expansion-matrix}
\end{align}
\end{subequations}
where, under the normalization condition $\bF^T\bC^{-1}\bF=\bI$, $\bPhi$ is the foreground projection matrix described in Section~\ref{sec:enumerating-assumptions}, given by $\bPhi=\bF\bF^T\bC^{-1}$. The calculations leading to Equations~\ref{eq:signal-channel-covariance-in-terms-of-expansion-matrix}~and~\ref{eq:signal-channel-mean-in-terms-of-expansion-matrix} are presented in Appendix~\ref{app:minimum-assumption-analysis-general-calculation}. Appendices~\ref{app:minimum-assumption-analysis-total-power}~and~\ref{app:minimum-assumption-analysis-full-stokes} give the forms of $\bDelta^{(21)}$ and $\bgamma^{(21)}$ for the specific cases of $N_s$ total power spectra, where the signal expansion matrix is $\bPsi=\begin{bmatrix} \bI & \bI & \cdots & \bI \end{bmatrix}^T$, and $4N_s$ spectra of all 4 Stokes parameters, where the expansion matrix is $\bPsi=\begin{bmatrix} \bI & \bzero & \bzero & \bzero & \cdots & \bI & \bzero & \bzero & \bzero \end{bmatrix}$.\footnote{For the sake of simplicity, results using the MAA with polarization are not shown in this paper, but the equations of Appendix~\ref{app:minimum-assumption-analysis-full-stokes} will prove useful in future work.}

\begin{figure}
  \centering
  \includegraphics[width=0.48\textwidth]{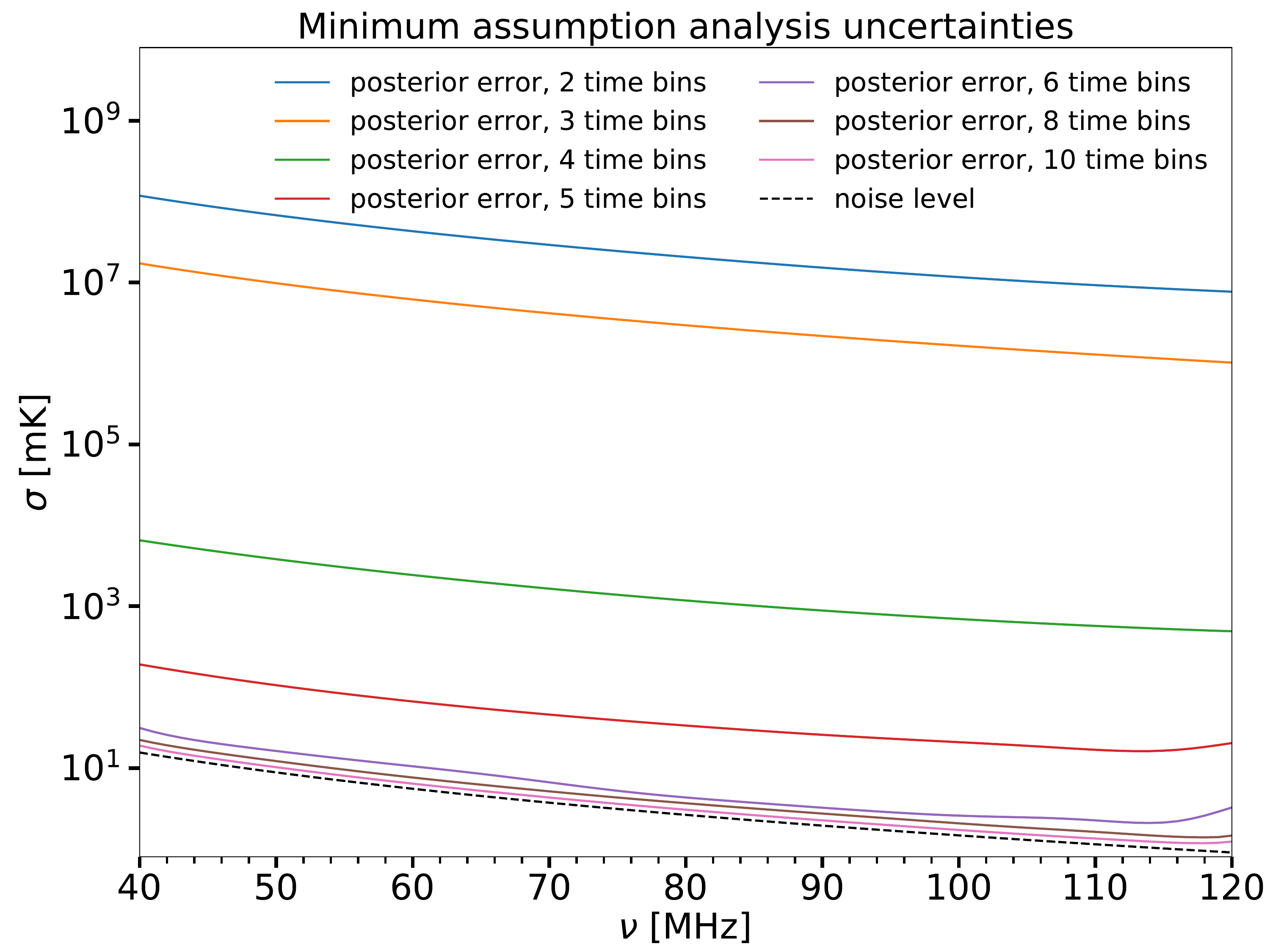}
  \caption{1$\sigma$ uncertainty levels under the minimum assumption analysis (MAA) with varying number of time bins. The dashed line represents the noise level on the signal. We assume an integration time of 100 hours split evenly across all time bins.} \label{fig:minimum-assumption-analysis-results}
\end{figure}

\begin{figure}
  \centering
  ~\includegraphics[width=0.47\textwidth]{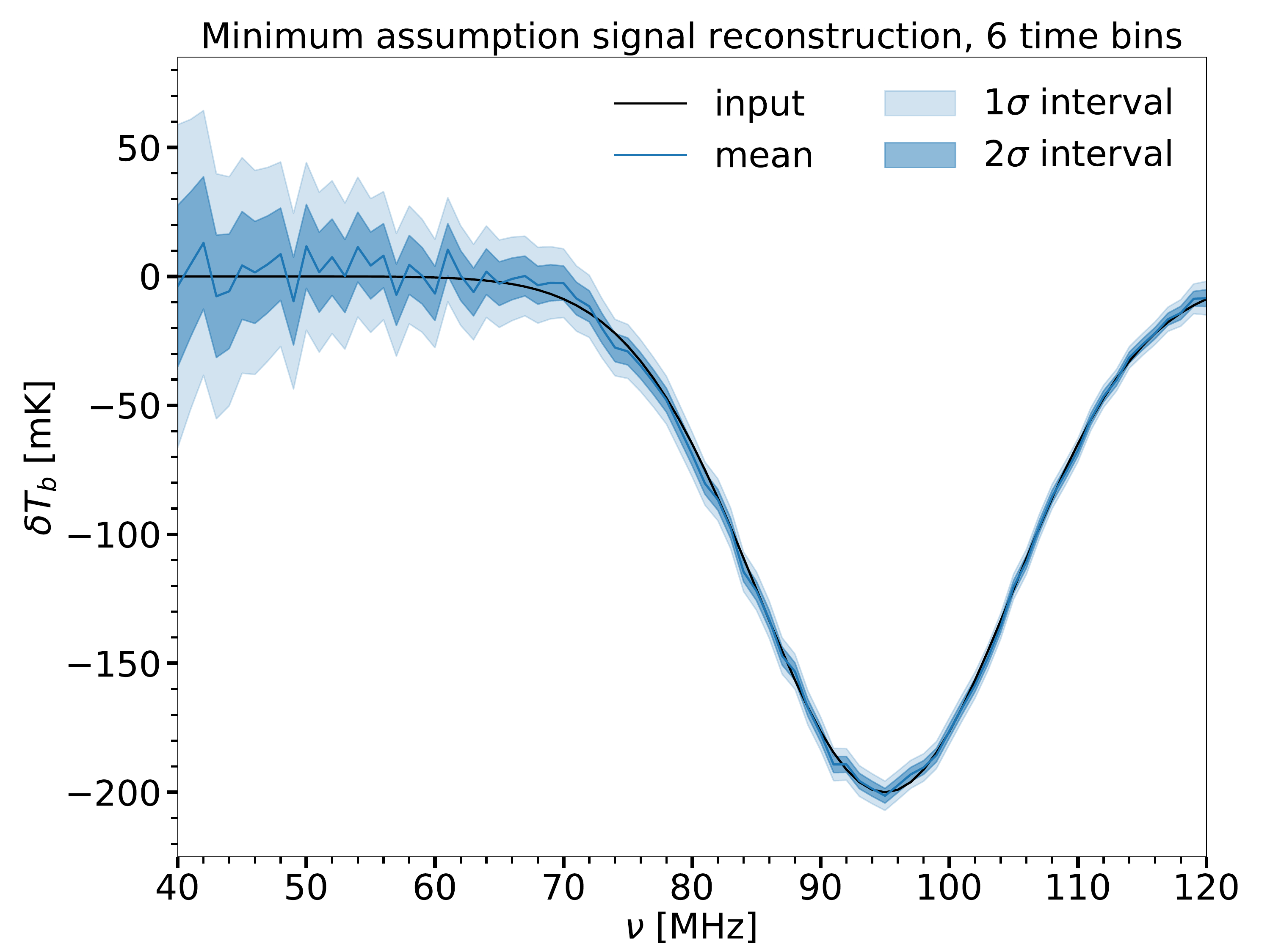}
  \includegraphics[width=0.48\textwidth]{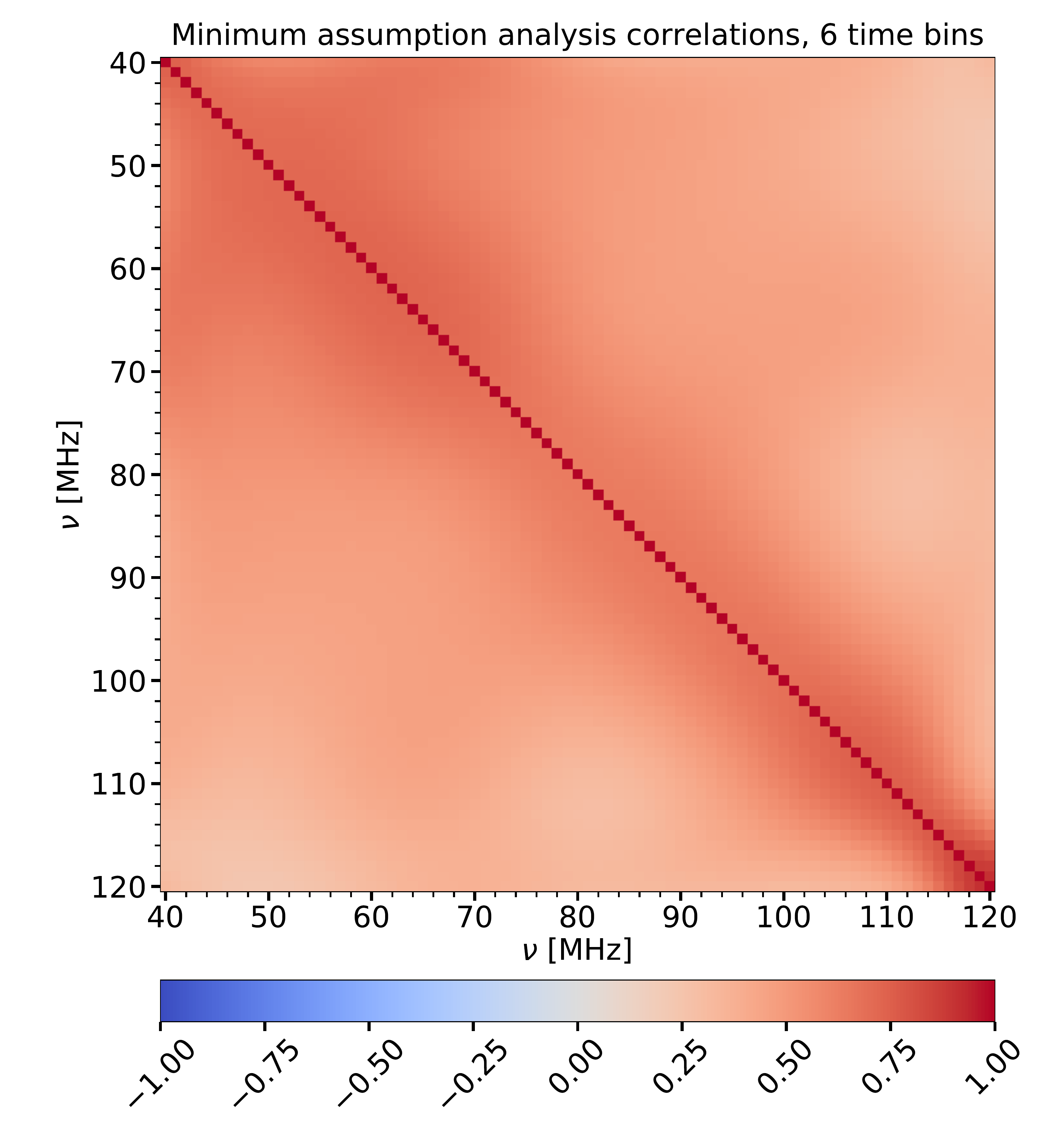}
  \caption{\textit{Top}: Example signal reconstruction for the MAA using 6 time bins. While the case shown is a Gaussian in frequency, note that the MAA can be used in the same manner obtaining equal errors to measure any possible signal spectrum. The black line shows the input signal. The blue line shows the channel mean of the reconstruction, given by $\bgamma^{(21)}$ in Equation~\ref{eq:signal-channel-mean-in-terms-of-expansion-matrix}, and the intervals show 1 and 2 times the square root of the diagonal elements of $\bDelta^{(21)}$ from Equation~\ref{eq:signal-channel-covariance-in-terms-of-expansion-matrix}. Because the beam-weighted foreground model (derived from the training set) fits the given beam-weighted foreground case (which was taken from the training set), these intervals correspond to 68\% and 95\% confidence levels. \textit{Bottom}: Frequency correlation matrix defined in Equation~\ref{eq:correlation-matrix}. Clearly, the foreground filter generated by this training set induces positive correlations between all frequencies of the 21-cm signal estimate. As in Figure~\ref{fig:minimum-assumption-analysis-results}, both panels assume an integration time of 100 hours split evenly across all time bins.} \label{fig:confidence-band-and-frequency_correlations}
\end{figure}

\subsection{Simulation details} \label{sec:minimum-assumption-analysis-simulation-details}

To test the MAA, we apply it using a foreground training set that is made identically to that used in \cite{Tauscher:20}, which we briefly review here.~Like the simulations from Section~\ref{sec:edges-mock-simulations-and-analysis}, the foreground map used in this training set was the Haslam map scaled down in frequency using a spatially constant spectral index of -2.5. The pointing of the antenna is set by the EDGES latitude and the sources and beam below the horizon are set to zero as in Section~\ref{sec:edges-mock-simulations-and-analysis}. The beams used have FWHMs specified by a distribution of quadratic functions given by $\alpha(\nu)=\sum_{k=0}^2b_kL_k\left(\frac{\nu - (80\text{ MHz})}{40\text{ MHz}}\right)$ where $L_k$ are the $k^{\text{th}}$ order Legendre polynomials and the $b_k$'s are normal with means $\mu_0=70^\circ$, $\mu_1=-20^\circ$, and $\mu_2=0^\circ$ and standard deviations $\sigma_0=10^\circ$, $\sigma_1=5^\circ$, and $\sigma_2=5^\circ$.

The day is split into 100 LST chunks,\footnote{In each of these $\sim$15 minute chunks, the foreground is smeared as shown in Figure~\ref{fig:edges-simulation-foreground}.} which are then averaged to the number of bins used for the given analysis. For example, in the 5 time bins case, LST chunks 1-20, 21-40, 41-60, 61-80, and 81-100 are used to generate the 5 time bins. If the number of time bins used does not divide 100, then the 100 chunks are reduced to the largest multiple of the number of bins less than 100 before binning.

\subsection{MAA results}

The results of applying MAA using the training set described in Section~\ref{sec:minimum-assumption-analysis-simulation-details} are shown in Figures~\ref{fig:minimum-assumption-analysis-results}~and~\ref{fig:confidence-band-and-frequency_correlations}. Figure~\ref{fig:minimum-assumption-analysis-results} shows the $1\sigma$ uncertainty levels at each frequency using different numbers of total power spectra in the training set.~These uncertainties are very large for small numbers of bins (indeed, they diverge when only one time bin is used), but approach the noise level for large numbers of bins.~Note that different training sets will lead to different results and that confidence levels of $68\%$, $95\%$, and $99.7\%$ only correspond to $1\sigma$, $2\sigma$, and $3\sigma$ when the training set adequately describes the beam-weighted foreground. If the training set is not adequate, any given percentage confidence interval will be wider than expected based on the $\sigma$-level. Appendix~\ref{app:minimum-assumption-analysis-signal-bias-statistic} derives the so-called signal bias statistic, $\varepsilon$, from \cite{Tauscher:18} in the MAA case when the foreground training set is imperfect. This statistic connects percentage confidence to the $\sigma$-level at which the signal is inside the interval in a root-mean-square sense.

The top panel of Figure~\ref{fig:confidence-band-and-frequency_correlations} shows an example reconstruction found when applying the MAA with 6 time bins to the sum of a beam-weighted foreground curve from the training set and an additional signal component that is Gaussian in frequency. The two intervals show the $1\sigma$ and $2\sigma$ confidence levels at each individual frequency, which are 1 and 2 times the square roots of the diagonal elements of $\bDelta^{(21)}$, respectively. Note, however, that even though a Gaussian signal is used in this example, as shown in the figure, the method would work equally well, providing the same errors, with any conceivable global 21-cm signal.

The bottom panel of Figure~\ref{fig:confidence-band-and-frequency_correlations} shows the frequency correlation matrix, $\bLambda^{(21)}$, which is a normalized form of the covariance matrix, $\bDelta^{(21)}$. The correlation matrix is designed to contain only values between 1 and -1, which occur at perfect correlation and anti-correlation, respectively. It is defined through $\Lambda^{(21)}_{ij}=\text{Corr}[\gamma_i,\gamma_j]=\text{Cov}[\gamma_i,\gamma_j]/\sqrt{\text{Var}[\gamma_i]\,\text{Var}[\gamma_j]}$, which can also be written as
\begin{equation}
  \bLambda^{(21)} = \left(\bGamma^{(21)}\right)^{-1/2} \bDelta^{(21)} \left(\bGamma^{(21)}\right)^{-1/2}, \label{eq:correlation-matrix}
\end{equation}
where $\bGamma^{(21)}$ has the same diagonal elements as $\bDelta^{(21)}$ but has no nonzero off-diagonal elements. By construction, all diagonal elements of $\bLambda^{(21)}$ are equal to 1. It is clear from Figure~\ref{fig:confidence-band-and-frequency_correlations} that all of the frequency channels are positively correlated with each other in this particular case.

To completely utilize the power of the MAA, the full covariance matrix, which would include such correlations, should be used when defining a likelihood function to extend MAA results to constrain any given signal model.

\section{Discussion} \label{sec:discussion}

\subsection{EDGES-like analysis} \label{sec:edges-like-discussion}

\subsubsection{Beam chromaticity distortions} \label{sec:beam-chromaticity-distortions}

The fits in Section~\ref{sec:edges-mock-simulations-and-analysis} (Figure~\ref{fig:edges-simulations}) show that false troughs can appear when fitting a single spectrum. Some may argue that troughs cannot be created by the foreground because the spectral dependence of the foreground does not allow it. However, one needs to keep in mind that what is measured is not the intrinsic spectral structure of the foreground radiation, but rather the spectral structure of the beam-weighted foreground. Since the angular structure of the beam weighting changes from frequency to frequency the foreground component of the measured spectrum is composed of different ratios of each direction's radiation at each frequency.

This means that a linear model that fits the intrinsic foreground spectrum of each pixel, like the model from Equation~\ref{eq:edges-foreground-model}, will not necessarily fit the beam-weighted foreground. To illustrate this, suppose that every sky direction's intrinsic foreground spectrum can be fit by a model using a basis matrix $\bF_0$, i.e. $\bT(\theta,\phi)=\bF_0\bx(\theta,\phi)$ where the $k^{\text{th}}$ element of $\bT(\theta,\phi)$ is the foreground spectrum at the $k^{\text{th}}$ frequency. If the beam is achromatic, then the beam-weighted foreground is given by
\begin{subequations}
\begin{align}
    \by^{(\fg)}_{\text{achromatic}} &= \int B(\theta,\phi)\ \bT(\theta,\phi)\ d\Omega, \\
    &= \bF_0\int B(\theta,\phi)\ \bx(\theta,\phi)\ d\Omega,
\end{align}
\end{subequations}
where $B(\theta,\phi)$ is the beam pattern at every frequency that satisfies $\int B(\theta,\phi)\ d\Omega=1$. This means that a linear model that fits $\bT(\theta,\phi)$ for every angle also fits the beam-weighted foreground $\by^{(\fg)}_{\text{achromatic}}$, i.e.~the foreground model residual is $\bdelta_{\text{achromatic}}=(\bI-\bPhi)\by^{(\fg)}_{\text{achromatic}}=\bzero$. In the case of beam chromaticity, the beam is a diagonal matrix with the diagonal entries being the beam patterns at each frequency, satisfying $
\int\bB(\theta,\phi)\ d\Omega = \bI$. In this case, the beam-weighted foreground is
\begin{subequations}
\begin{align}
    \by^{(\fg)}_{\text{chromatic}} &= \int \bB(\theta,\phi)\ \bT(\theta,\phi)\ d\Omega \\
    &= \int \bB(\theta,\phi)\ \bF_0\bx(\theta,\phi)\ d\Omega
\end{align}
\end{subequations}
and the residual
\begin{equation}
    \bdelta_{\text{chromatic}} = \int (\bI-\bPhi)\bB(\theta,\phi)\ \bF_0\bx(\theta,\phi)\ d\Omega
\end{equation}
is not necessarily zero,\footnote{The only situation in which $\bdelta^{\text{chromatic}}$ is necessarily zero for all beams, $\bB(\theta,\phi)$, occurs when $\bI=\bPhi$. But, this implies that projecting into the span of the foreground basis has no effect, which can only happen if the foreground basis is complete. In this case, the signal model will be degenerate with the foreground model and therefore cannot be extracted.} leading to a possible failure of the assumption that the foreground model is adequate (Assumption~\ref{item:foreground-assumption}). Therefore, while troughs are unlikely to appear in intrinsic foreground spectra, they may well appear when beam-weighted foregrounds are fit with a model of a trough and an intrinsic foreground spectrum model simultaneously.

\subsubsection{EDGES beam chromaticity factor}

In several of their works \citep[see, e.g.,][]{Bowman:18, Monsalve:18,Mozdzen:19}, the EDGES team uses what they term the ``beam chromaticity factor'' \citep[see Equation 14 of][]{Monsalve:17b}, defined at each LST through
\begin{equation}
    C(\nu) = \frac{\int B_{\text{ref}}(\nu,\theta,\phi)\ T_{\text{ref}}(\nu,\theta,\phi)\ d\Omega}{\int B_{\text{ref}}(\nu_{\text{ref}},\theta,\phi)\ T_{\text{ref}}(\nu,\theta,\phi)\ d\Omega},
\end{equation}
where the LST dependence comes from how the temperature map is rotated with respect to the beam and $B_{\text{ref}}(\nu,\theta,\phi)$ and $T_{\text{ref}}(\nu,\theta,\phi)$ are the assumed forms of the beam and foreground. The goal of this factor is to make the spectrum appear as if it was measured with an achromatic beam (specifically, the beam at $\nu=\nu_{\text{ref}}$) so that, as discussed in Section~\ref{sec:beam-chromaticity-distortions}, basis vectors that fit the intrinsic foreground spectra can be used to fit the beam-weighted foreground spectrum. If we denote a spectrum measured at a single LST by $y(\nu)$, then the EDGES beam calibration forms a corrected spectrum, $y^\prime(\nu)=y(\nu)/C(\nu)$. Neglecting noise and possible receiver errors for the sake of clarity, we can express the measured beam-weighted foreground spectrum through
\begin{equation}
    y(\nu) = \int B(\nu,\theta,\phi)\ T(\nu,\theta,\phi)\ d\Omega,
\end{equation}
where $B(\nu,\theta,\phi)$ and $T(\nu,\theta,\phi)$ are the unknown forms of the true beam and foreground. With these definitions, the corrected spectrum is given by
\begin{multline}
  y^\prime(\nu) = \int B_{\text{ref}}(\nu_{\text{ref}},\theta,\phi)\ T_{\text{ref}}(\nu,\theta,\phi)\ d\Omega \\ \times \frac{\int B(\nu,\theta,\phi)\ T(\nu,\theta,\phi)\ d\Omega}{\int B_{\text{ref}}(\nu,\theta,\phi)\ T_{\text{ref}}(\nu,\theta,\phi)\ d\Omega}. \label{eq:beam-weighted-corrected-spectrum}
\end{multline}
The intention is for the fraction on the second line of Equation~\ref{eq:beam-weighted-corrected-spectrum} to be 1 so that $y^\prime(\nu)$ is simply the reference foreground weighted by the reference beam at the reference frequency. However, in order to assume that the fraction is 1, one must assume that $B_\text{ref}(\nu,\theta,\phi)=B(\nu,\theta,\phi)$ and $T_{\text{ref}}(\nu,\theta,\phi)=T(\nu,\theta,\phi)$; but, this is not a reasonable assumption as foreground and beam models are likely imperfect.\footnote{If the reference beam and foreground were equal to the true beam and foreground, then one would know the exact beam-weighted foreground and should simply subtract it from the measured spectrum, leaving only noise and the 21-cm signal.} The beam chromaticity factor calibration method is therefore prone to introduce significant biases. On the contrary, our method of deriving modes describing the beam-weighted foreground from a large sample of simulations and observations allows uncertainties in both the beam and foreground to be included in the analysis. However, as will be elaborated on in Section~\ref{sec:minimum-assumption-analysis-discussion}, it is important to note that our method is contingent on the ability of the simulations and observations to accurately describe the data being analyzed. SVD can only provide eigenmodes in the space spanned by the training set curves.

\subsubsection{Toy galaxy model}
\label{sec:toy-galaxy-model}

In Section~\ref{sec:edges-mock-simulations-and-analysis}, in addition to the Haslam 408 MHz map, we also tested a toy model of galactic emission described by
\begin{multline}
  T_{\text{toy}}(\nu,\theta,\phi) = \left(25\text{ K}\right)\ \left(\frac{\nu}{408\text{ MHz}}\right)^{-2.5} \\ \times \left\{1 + \left[9 + 2\cos{\phi}\right]2^{-\left[\left(\theta - \frac{\pi}{2}\right)/\left(\frac{\pi}{45}\right)\right]^2}\right\}.
\end{multline}
This is a model that treats the Galactic plane as a $8^\circ$ FWHM Gaussian in colatitude, $\theta$, with a maximum given by 300 K at the Galactic center and 200 K at the anticenter and asymptotes to 25 K away from the plane. This map is then scaled by $[\nu/(408\text{ MHz})]^{-2.5}$. Due to the fact that the false troughs remain after this change (see e.g. the dashed green line in the lower left panel of Figure~\ref{fig:edges-simulations}), we infer that the chromaticity of the quadratic beams used in Section~\ref{sec:edges-mock-simulations-and-analysis} interacts with the galactic plane as seen from the EDGES latitude to generate residuals that are better fit with a flattened Gaussian plus foreground model than the foreground model alone, which could lead to analyses falsely concluding that there are troughs in the sky-averaged radio spectrum.

Since the false troughs found in Section~\ref{sec:edges-mock-simulations-and-analysis} are fits to aspects of the beam-weighted foreground, we urge EDGES to perform the same experiment and analysis from the northern hemisphere where the galaxy behaves very differently in the sky than from the southern hemisphere. Because of the different orientation of the galactic plane, the distortions caused by chromatic beams in our simulations are more easily fit by foreground models like that in Equation~\ref{eq:edges-foreground-model} at northern latitudes than southern latitudes.

\subsubsection{Exploring parameter distributions}

The natural endpoint of the EDGES-like analysis is not a maximum likelihood fit like the one shown in Figure 1 of \cite{Bowman:18}, on which Figure~\ref{fig:edges-simulations} was based. Instead, one wishes to explore the parameter distribution using a sampling method like Markov Chain Monte Carlo (MCMC) exploration. However, the resulting distribution is only meaningful if the pieces of the model (i.e. foreground and signal models) accurately describe their respective components. In fact, parameter distributions from global 21-cm signal fits to observations are generated by a complex combination of the following four factors: 1) the foreground model parameterization; 2) the bias of the foreground model; 3) the signal model parameterization; 4) the bias of the signal model.

Ideally, factors 2 and 4 are unimportant because the foreground and signal models can fit the true foreground and signal to within the noise level. However, it is impossible to fully verify that this is the case in practice; so, their effects must be considered. Section~\ref{sec:edges-mock-simulations-and-analysis} is not meant to explain the exact form and parameter distribution of the trough observed by EDGES, but instead to show that generic polynomial foreground models like those used by EDGES do not account for beam chromaticity and may lead to untrustable results. Essentially, we are pointing out that factor 2 is likely affecting the EDGES analysis.

In addition to generating bias in fits, residuals from fits with inaccurate foreground models may have led EDGES to choose the unjustified flattened Gaussian signal model, possibly furthering the signal bias of factor 4 (with respect to a more conventional signal model) silently while significantly reducing overall residuals.\footnote{If true, this would be a case of factor 2 affecting factor 3. For robustness reasons, it is important to justify the signal model in a way external to the observations themselves.}

Due to the complex interactions between the four factors in generating a posterior distribution, MCMC exploration of distributions is outside the scope of this paper and is left for an extended examination in future work.

\subsection{Minimum assumption analysis} \label{sec:minimum-assumption-analysis-discussion}

\subsubsection{Advantages and limitations}

While moving the experiment to a different location could provide evidence that the reported trough is a product of an inadequate foreground model, this does not solve the underlying problem---that the foreground model does not account for beam chromaticity.

The MAA proposed in Section~\ref{sec:minimum-assumption-analysis} avoids this problem by making a basis specific to the given antenna by performing SVD on a training set of simulations made with that antenna's beam (see Section~\ref{sec:choosing-foreground-basis}).

In addition, the MAA allows any possible global signal, avoiding unwanted bias from generating a signal model that interacts with the foreground model bias to produce false results. In fact, under the assumption that the foreground model generated by SVD is adequate, the MAA signal mean $\bgamma^{(21)}$ and covariance $\bDelta^{(21)}$ constitute a direct measurement not of a signal model, but of the signal itself. In fact, since all degrees of freedom are retained, a physical model can be fit directly to the constraints implied by $\bgamma^{(21)}$ and $\bDelta^{(21)}$ (such as those shown in the bottom panel of Figure~\ref{fig:minimum-assumption-analysis-results}).

Note that, for simplicity, the cosmic microwave background (CMB) is not included in the training set used in Section~\ref{sec:minimum-assumption-analysis}, which merely uses the Haslam map scaled with a spectral index of -2.5. If it was included (as it should be when analyzing data from a real experiment),\footnote{More precisely, the CMB should be subtracted from the foreground model used to generate the training set so that it can be found by the signal fitting and subtracted after the fact.} then the signal mean $\bgamma^{(21)}$ would include a 2.725 K flat spectrum component which would need to be subtracted out.

Another important note about the MAA is that it is predicated upon the beam-weighted foreground training set provided to it. The results may change if the training set is changed, even if the data being analyzed remain the same. The training set used in Section~\ref{sec:minimum-assumption-analysis} contained a wide variety of beam FWHMs but only one intrinsic foreground map. A different training set built using many intrinsic foreground maps and one beam might produce errors that differ significantly from the ones presented in Figure~\ref{fig:minimum-assumption-analysis-results}. However, even as different training sets would lead to different levels of precision, the accuracy of the MAA should remain steady as the training set changes as long as the true beam-weighted foreground is encompassed by the SVD eigenmodes of the training sets.\footnote{Here, we use precision to refer to the size of uncertainties, which is known even when the true 21-cm signal is unknown and accuracy to refer to the size of the difference between the signal reconstruction and the true signal with respect to the size of the uncertainties.} One aspect of this effect is that some training sets will require more modes to be included in the foreground model than others, forcing one to degrade precision in order to retain accuracy.

\subsubsection{Extension to motion induced dipole} \label{sec:motion-induced-dipole}

\cite{Slozar:17} pointed out that, much like with the Cosmic Microwave Background (CMB), Doppler shifting of the 21-cm background induces a dipole component that is related to the monopole. In particular, the dipole component, $\delta T_b^{(\text{dip})}$, of the 21-cm signal is related to the monopole spectrum, $\delta T_b^{(\text{mon})}$, through\footnote{Note that this assumes that there is no intrinsic (i.e. non-motion-induced) dipole of the 21-cm signal.}
\begin{equation}
  \delta T_b^{(\text{dip})}(\nu,\psi) = \beta\cos{\psi}\left(1-\nu\frac{d}{d\nu}\right)\delta T_b^{(\text{mon})}(\nu), \label{eq:motion-induced-dipole}
\end{equation}
where $\beta$ is the magnitude of our velocity with respect to the CMB as a fraction of the speed of light and $\psi$ is the angle between that velocity and the line of sight. Based on CMB observations, $\beta\approx 1.2\times 10^{-3}$. \cite{Deshpande:18} suggested that Equation~\ref{eq:motion-induced-dipole} should be considered an essential qualifying test of any measurement of the global signal. Due to the fact that (by Equation~\ref{eq:motion-induced-dipole}), $\delta T_b^{(\text{dip})}$ is linear in $\delta T_b^{(\text{mon})}$, it could conceivably be included in the MAA in order for any fit to pass this test by default, essentially extending the minimal signal assumption introduced in Section~\ref{sec:minimum-signal-assumption}. To do so, we note that the sum of the monopole and dipole components is
\begin{equation}
  \delta T_b^{(\text{mon}+\text{dip})}(\nu,\psi) = \left[1 + \beta\cos{\psi}\left(1-\nu\frac{d}{d\nu}\right)\right]s(\nu),
\end{equation}
where $s(\nu)=\delta T_b^{(\text{mon})}(\nu)$. To determine how this would appear to a single antenna experiment, we must multiply by the beam and integrate over all angles. Defining $t_k(\nu)\equiv\int B_k(\nu,\theta,\phi)\ \delta T_b^{(\text{mon}+\text{dip})}(\nu,\psi)\ d\Omega$ as the signature of the 21-cm signal in the spectrum measured at the $k^{\text{th}}$ time period (or, equivalently, pointing angle), it is clear that
\begin{equation}
  t_k(\nu) = \left[1+\beta a_k(\nu)\left(1-\nu\frac{d}{d\nu}\right)\right]s(\nu),
\end{equation}
where $a_k(\nu)=\int B_k(\nu,\theta,\phi)\ \cos{\psi}\ d\Omega$.\footnote{Note that $a_k(\nu)$ will not be known exactly because the beam will not be known exactly. However, since the dipole component will be on the order of 10 mK \citep[see][]{Slozar:17} and $a_k(\nu)$ will be of order unity, the expected imprecision levels in the beam, which should be at the sub-percent level, should not impact the results significantly.} This can be written in matrix notation where frequency channels correspond to the elements of vectors. In this way, the signature of the 21-cm signal in the $k^{\text{th}}$ spectrum is
\begin{equation}
  \bt_k = \left[\bI + \beta\bA_k(\bI-\bN\bD)\right]\bs,
\end{equation}
where $\bA_k$ is the diagonal matrix with nonzero elements given by $\ba_k$, $\bN$ is the diagonal matrix with the frequencies of the data channels on the diagonal, and $\bD$ is a derivative matrix (see Appendix~\ref{app:minimum-assumption-analysis-derivative-for-motion-induced-dipole} for subtleties involved in taking derivatives with a matrix). Now, combining all $N_s$ spectra into one vector leaves $\bt=\bPsi\bs$ where
\begin{equation}
  \bPsi=\begin{bmatrix} \bI \\ \bI \\ \vdots \\ \bI \end{bmatrix}  + \beta\begin{bmatrix} \bA_1 \\ \bA_2 \\ \vdots \\ \bA_{N_s} \end{bmatrix} (\bI-\bN\bD).
\end{equation}
To extend the MAA to include the motion-induced dipole, one must merely plug this $\bPsi$ into Equations~\ref{eq:signal-channel-covariance-in-terms-of-expansion-matrix}~and~\ref{eq:signal-channel-mean-in-terms-of-expansion-matrix}. The effect of including the motion-induced dipole on the results of the MAA is left for future work.

\subsection{Between the two extremes}

In past work \citep{Tauscher:18,Rapetti:19,Tauscher:20}, we laid out an SVD-based pipeline for extracting the 21-cm global signal from the large beam-weighted foregrounds. That analysis is identical in its foreground basis generation to the MAA from Section~\ref{sec:minimum-assumption-analysis}. However, it differs in that the signal is restricted to a specific model, either a physically motivated one or one created from performing SVD on a signal training set. So, in a sense, the pipeline described in those works is between the two extremes discussed in this paper of strong assumptions (EDGES-like analysis) and robust assumptions (MAA).

\section{Conclusions} \label{sec:conclusions}

In this paper, we formulated a list of general assumptions for global 21-cm signal analyses. These include assumptions about the calibration of the instrument (Assumption~\ref{item:calibration-assumption}), the distribution of the noise (Assumption~\ref{item:noise-assumption}), the adequacy of the foreground model (Assumption~\ref{item:foreground-assumption}), and the form of the signal (Assumption~\ref{item:signal-assumption}). We then contrasted two different specific forms of these assumptions, an EDGES-like analysis and a new, so-called minimum assumption analysis (MAA).

The EDGES-like analysis is performed on single spectra with a polynomial-based foreground model and a flattened Gaussian signal model. We presented fits of simulations using zero signal and the Haslam map (as seen from the EDGES latitude) scaled with a constant spectral index and weighted by four different Gaussian beams with quadratic FWHMs. Two of these fits resulted in large flattened Gaussian troughs near 78 MHz, like the one reported by the EDGES team. These troughs also appeared when the Haslam map was replaced by a toy model of galactic emission that uses a 25 K background temperature (at 408 MHz) and models the galactic plane as a Gaussian in latitude with a peak of 300 K at the galactic center and 200 K at the anticenter, showing that the interaction of the beam with the Galactic plane introduces troughs. 

This vulnerability to false troughs is due to the inadequacy of the foreground model, which does not account for beam chromaticity. While the only real solution to this problem is to modify the analysis technique, steps such as moving the experiment to the northern hemisphere, where the galaxy appears very differently, should at least modify (and potentially decrease) any false troughs that are found.

The MAA, on the other hand, is performed on many time-binned spectra \citep[see also][]{Tauscher:20} instead of one spectrum. Also, instead of assuming a specific model for the signal, it allows for any possible spectrum that is the same across each time bin. In addition, the foreground model accounts for beam chromaticity because the basis vectors are taken from applying Singular Value Decomposition (SVD) to a training set of foregrounds weighted by the beam of the specific antenna being used \citep{Tauscher:18}. Given the beam-weighted foreground training set employed, we found that, under these assumptions, any signal can be measured with uncertainties within an order of magnitude of the noise level.

While the MAA could be considered the most robust, conservative form of the assumptions specified in Section~\ref{sec:enumerating-assumptions}, if there are well motivated theoretical models for the signal, the pipeline we laid out in \cite{Tauscher:18}, \cite{Rapetti:19}, and \cite{Tauscher:20} can be applied. That method uses training sets for both the beam-weighted foreground and the global signal to create models for each of them. Ultimately, experimenters  can decide if a theoretical model of the signal (not just based on residuals from the data with respect to the intrinsic foreground model, as in the case of the flattened Gaussian) is rigorous enough to be explored using our full pipeline, for stronger constraints. However, given the current theoretical uncertainties, if the MAA is possible for the selected beam-weighted foreground model, we recommend to start with this analysis before selecting a physical model of the signal because it allows for any signal---even those which are unexpected---and may guide the model selection procedure.

\acknowledgments{We thank Neil Bassett and Joshua Hibbard for useful discussions during the development of this study. We also appreciate detailed discussions with Judd Bowman and Raul Monsolve about the EDGES data analysis approach.~This work is directly supported by the NASA Solar System Exploration Virtual Institute cooperative agreement 80ARC017M0006.}

%\facility{facility ID}
%\facilities{facility ID, facility ID, facility ID} 
%\software{Numpy}

\appendix

\section{Minimum assumption analysis general calculation} \label{app:minimum-assumption-analysis-general-calculation}

In this appendix, we will compute the minimum assumption maximum likelihood for signal reconstruction and the uncertainties on that reconstruction when the signal expansion matrix is $\bPsi$ (i.e. the signal $\by^{(21)}$ appears in the data as $\bPsi\by^{(21)}$), the noise covariance is $\bC$, and the foreground basis matrix is $\bF$.

The signal assumption may be implemented by assuming $\by^{(21)}=\bA\bx^{(21)}$ for an invertible\footnote{Note that invertible matrices are square, meaning that there are as many parameters in $\bx^{(21)}$ as there are frequencies in $\by^{(21)}$.} basis matrix $\bA$.\footnote{$\bA$ could be assumed to be the identity matrix, but the analytical calculations can be completed more easily if the signal basis matrix is normalized.} The model for the full data is
\begin{subequations}
\begin{align}
  \bmM(\bx^{(\fg)},\bx^{(21)}) &= \bmM^{(\fg)}(\bx^{(\fg)}) + \bmM^{(21)}(\bx^{(21)}) \\
  &= \bF\bx^{(\fg)} + \bPsi\bA\bx^{(21)},
\end{align}
\end{subequations}
This can also be written $\bmM(\bx) = \bG\bx$, where $\bG = \begin{bmatrix} \bF & \bPsi\bA \end{bmatrix}$, $\bx=\begin{bmatrix} \bx^{(\fg)} \\ \bx^{(21)} \end{bmatrix}$, and $\bF$ is the foreground basis that applies to all of the spectra simultaneously, as opposed to the single spectrum basis laid out in Section~\ref{sec:enumerating-assumptions}. The likelihood function is therefore
\begin{equation}
  \mL(\bx) \propto e^{-(\by-\bG\bx)^T\bC^{-1}(\by-\bG\bx)/2}.
\end{equation}
The maximum likelihood value of $\bx$, $\bxi$, and its covariance, $\bS$, are given by
\begin{equation}
    \bxi = \bS\bG^T\bC^{-1}\by
    \ \ \text{ and } \ \   \bS = (\bG^T\bC^{-1}\bG)^{-1}.
\end{equation}
These are easiest to calculate when $\bF$ and $\bPsi\bA$ are normalized through $\bF^T\bC^{-1}\bF=\bI$ and $\bA^T\bPsi^T\bC^{-1}\bPsi\bA=\bI$. Under these conditions, the covariance of the signal parameters is
\begin{equation}
  \bS^{(21)} = (\bI - \bA^T\bPsi^T\bC^{-1}\bPhi\bPsi\bA)^{-1},
\end{equation}
where $\bPhi$ is the projection matrix described in Section~\ref{sec:enumerating-assumptions}, which is given by $\bPhi=\bF\bF^T\bC^{-1}$ under the current normalization conditions. This means that the channel covariance of the signal distribution is $\bDelta^{(21)} = \bA\bS^{(21)}\bA^T$, which can be written as
\begin{equation}
  \bDelta^{(21)} = \left[\left(\bA\bA^T\right)^{-1} - \bPsi^T\bC^{-1}\bPhi\bPsi\right]^{-1}. \label{eq:minimum-assumption-signal-channel-covariance}
\end{equation}
The mean of the signal parameters is
\begin{equation}
  \bxi^{(21)} = \bS^{(21)}\bA^T\bPsi^T\bC^{-1}\left(\bI-\bPhi\right)\by
\end{equation}
and the channel mean of the signal distribution is $\bgamma^{(21)}=\bA\bxi^{(21)}$, or
\begin{equation}
    \bgamma^{(21)} = \bDelta^{(21)}\bPsi^T\bC^{-1}(\bI-\bPhi)\by. \label{eq:intermediate-mean-in-terms-of-expansion-matrix}
\end{equation}
To satisfy our normalization condition, the signal basis matrix $\bA$ must satisfy $\bA^T\bPsi^T\bC^{-1}\bPsi\bA = \bI$. Multiplying on the left by $(\bA\bA^T)^{-1}\bA$ and on the right by $\bA^{-1}$, we find that $(\bA\bA^T)^{-1} = \bPsi^T\bC^{-1}\bPsi$. Therefore, Equation~\ref{eq:minimum-assumption-signal-channel-covariance} becomes
\begin{equation}
  \bDelta^{(21)} = \left[\bPsi^T\bC^{-1}(\bI - \bPhi)\bPsi\right]^{-1}. \label{eq:final-covariance-in-terms-of-expansion-matrix}
\end{equation}
This means that if any vector in the column space of $\bPsi$, i.e. any vector of the form $\bPsi\bz=\begin{bmatrix} \bz^T & \bz^T & \cdots & \bz^T \end{bmatrix}^T$, can be represented by the foreground vectors, then the uncertainties are infinite. Plugging Equation~\ref{eq:final-covariance-in-terms-of-expansion-matrix} into Equation~\ref{eq:intermediate-mean-in-terms-of-expansion-matrix}, we find that the signal channel mean is
\begin{equation}
  \bgamma^{(21)} = \left[\bPsi^T\bC^{-1}(\bI-\bPhi)\bPsi\right]^{-1}\bPsi^T\bC^{-1}(\bI-\bPhi)\by.
\end{equation}

\section{Minimum assumption analysis with only total power spectra} \label{app:minimum-assumption-analysis-total-power}

As mentioned in Section~\ref{sec:minimum-signal-assumption}, when there are $N_s$ spectra concatenated together in the data curve being fit, i.e. $\by=\begin{bmatrix} \by_1^T & \by_2^T & \cdots & \by_{N_s}^T \end{bmatrix}^T$,
the signal expansion matrix $\bPsi$ is given by $\bPsi = \begin{bmatrix} \bI & \bI & \cdots & \bI \end{bmatrix}^T$. Assuming the different spectra have independent noise, we can write the covariance $\bC$ in this case as
\begin{equation}
    \bC = \begin{bmatrix} \bC_1 & \bzero & \cdots & \bzero \\ \bzero & \bC_2 & \cdots & \bzero \\ \vdots & \vdots & \ddots & \vdots \\ \bzero & \bzero & \cdots & \bC_{N_s} \end{bmatrix}.
\end{equation}
We split $\bF$ into $N_s$ different components through $\bF=\begin{bmatrix}\bF_1^T&\bF_2^T&\cdots&\bF_{N_s}^T\end{bmatrix}^T$. This means that the normalization condition of the foreground basis matrix, $\bF^T\bC^{-1}\bF=\bI$, becomes $\sum_{k=1}^{N_s} \bF_k^T\bC_k^{-1}\bF_k = \bI$,
the foreground projection matrix is
\begin{equation}
  \bPhi = \begin{bmatrix} \bF_1\bF_1^T\bC_1^{-1} & \bF_1\bF_2^T\bC_2^{-1} & \cdots & \bF_1\bF_{N_s}^T\bC_{N_s}^{-1} \\ \bF_2\bF_1^T\bC_1^{-1} & \bF_2\bF_2^T\bC_2^{-1} & \cdots & \bF_2\bF_{N_s}^T\bC_{N_s}^{-1} \\ \vdots & \vdots & \ddots & \vdots \\ \bF_{N_s}\bF_1^T\bC_1^{-1} & \bF_{N_s}\bF_2^T\bC_2^{-1} & \cdots & \bF_{N_s}\bF_{N_s}^T\bC_{N_s}^{-1} \end{bmatrix},
\end{equation}
the signal channel covariance matrix is
%\begin{equation}
%  \bDelta^{(21)} = \left(\sum_{k=1}^{N_s}\bC_k^{-1} - \sum_{m=1}^{N_s}\sum_{n=1}^{N_s}\bC_m^{-1}\bF_m\bF_n^T\bC_n^{-1}\right)^{-1},
%\end{equation}
\begin{equation}
  \bDelta^{(21)} = \left[\sum_{n=1}^{N_s}\left(\bI - \sum_{m=1}^{N_s}\bC_m^{-1}\bF_m\bF_n^T\right)\bC_n^{-1}\right]^{-1}, \label{eq:signal-channel-covariance-as-sum}
\end{equation}
and the signal channel mean is
%\begin{multline}
%  \bgamma^{(21)} = \\ \bDelta^{(21)} \left(\sum_{k=1}^{N_s}\bC_k^{-1}\by_k - \sum_{m=1}^{N_s}\sum_{n=1}^{N_s}\bC_m^{-1}\bF_m\bF_n^T\bC_n^{-1}\by_n\right).
%\end{multline}
\begin{equation}
  \bgamma^{(21)} = \bDelta^{(21)} \left[\sum_{n=1}^{N_s}\left(\bI-\sum_{m=1}^{N_s}\bC_m^{-1}\bF_m\bF_n^T\right)\bC_n^{-1}\by_n\right].
\end{equation}

\section{Minimum assumption analysis with polarization} \label{app:minimum-assumption-analysis-full-stokes}

In this appendix, we will layout the form of the MAA when the data vector is the concatenation of $4N_b$ spectra containing all four Stokes parameters at $N_b$ time bins, i.e.
\begin{equation}
    \by = \begin{bmatrix} \by_{1,I}^T & \by_{1,Q}^T & \by_{1,U}^T & \by_{1,V}^T & \cdots & \by_{N_b,I}^T & \by_{N_b,Q}^T & \by_{N_b,U}^T & \by_{N_b,V}^T \end{bmatrix}^T.
\end{equation}
The signal appears only in the Stokes I spectra, so the signal expansion matrix is
\begin{equation}
    \bPsi = \begin{bmatrix} \bI & \bzero & \bzero & \bzero & \cdots & \bI & \bzero & \bzero & \bzero \end{bmatrix}^T.
\end{equation}
Since the four Stokes parameters of the $k^{\text{th}}$ time bin have roughly the same noise covariance,\footnote{If $Q/I$, $U/I$, and $V/I$ have magnitudes of order $x$, then the fractional difference between the noise levels on $I$, $Q$, $U$, and $V$ is of order $x^2$. See \cite{Tauscher:20} for exact calculations of Stokes parameters noise levels. For the purposes here, it is best just to assume all four Stokes parameters have the same noise level.} $\bC_k$, we can write the full noise covariance matrix as
\begin{equation}
  \bC = \begin{bmatrix} \bC_1 & \bzero & \bzero & \bzero & \cdots & \bzero & \bzero & \bzero & \bzero \\ \bzero & \bC_1 & \bzero & \bzero & \cdots & \bzero & \bzero & \bzero & \bzero \\ \bzero & \bzero & \bC_1 & \bzero & \cdots & \bzero & \bzero & \bzero & \bzero \\ \bzero & \bzero & \bzero & \bC_1 & \cdots & \bzero & \bzero & \bzero & \bzero \\ \vdots & \vdots & \vdots & \vdots & \ddots & \vdots & \vdots & \vdots & \vdots \\ \bzero & \bzero & \bzero & \bzero & \cdots & \bC_{N_b} & \bzero & \bzero & \bzero \\ \bzero & \bzero & \bzero & \bzero & \cdots & \bzero & \bC_{N_b} & \bzero & \bzero \\ \bzero & \bzero & \bzero & \bzero & \cdots & \bzero & \bzero & \bC_{N_b} & \bzero \\ \bzero & \bzero & \bzero & \bzero & \cdots& \bzero & \bzero & \bzero & \bC_{N_b}  \end{bmatrix}.
\end{equation}
In this case, the foreground basis returned by the SVD algorithm is given by
\begin{equation}
    \bF = \begin{bmatrix} \bF_{1,I}^T & \bF_{1,Q}^T & \bF_{1,U}^T & \bF_{1,V}^T & \cdots & \bF_{N_b,I}^T & \bF_{N_b,Q}^T & \bF_{N_b,U}^T & \bF_{N_b,V}^T \end{bmatrix}^T,
\end{equation}
which implies that the normalization convention for the foreground basis matrix is $\sum_{k=1}^{N_s}\sum_{S\in\{I,Q,U,V\}}\bF_{k,S}^T\bC_k^{-1}\bF_{k,S}=\bI$,
and the foreground projection matrix is
\begin{equation}
    \bPhi = \begin{bmatrix} \bF_{1,I}\bF_{1,I}^T\bC_1^{-1} & \cdots & \bF_{1,I}\bF_{1,V}^T\bC_1^{-1} & \cdots & \bF_{1,I}\bF_{N_b,I}^T\bC_{N_b}^{-1} & \cdots & \bF_{1,I}\bF_{N_b,V}^T\bC_{N_b}^{-1} \\ \vdots & \ddots & \vdots & \ddots & \vdots & \ddots & \vdots \\ \bF_{1,V}\bF_{1,I}^T\bC_1^{-1} & \cdots & \bF_{1,V}\bF_{1,V}^T\bC_1^{-1} & \cdots & \bF_{1,V}\bF_{N_b,I}^T\bC_{N_b}^{-1} & \cdots & \bF_{1,V}\bF_{N_b,V}^T\bC_{N_b}^{-1} \\ \vdots & \ddots & \vdots & \ddots & \vdots & \ddots & \vdots \\ \bF_{N_b,I}\bF_{1,I}^T\bC_1^{-1} & \cdots & \bF_{N_b,I}\bF_{1,V}^T\bC_1^{-1} & \cdots & \bF_{N_b,I}\bF_{N_b,I}^T\bC_{N_b}^{-1} & \cdots & \bF_{N_b,I}\bF_{N_b,V}^T\bC_{N_b}^{-1} \\ \vdots & \ddots & \vdots & \ddots & \vdots & \ddots & \vdots \\ \bF_{N_b,V}\bF_{1,I}^T\bC_1^{-1} & \cdots & \bF_{N_b,V}\bF_{1,V}^T\bC_1^{-1} & \cdots & \bF_{N_b,V}\bF_{N_b,I}^T\bC_{N_b}^{-1} & \cdots & \bF_{N_b,V}\bF_{N_b,V}^T\bC_{N_b}^{-1} \end{bmatrix}.
\end{equation}
Very similarly to Equation~\ref{eq:signal-channel-covariance-as-sum}, the signal channel covariance given by Equation~\ref{eq:signal-channel-covariance-in-terms-of-expansion-matrix} is
\begin{equation}
    \bDelta^{(21)} = \left\{\sum_{n=1}^{N_b} \left[ \bI - \left(\sum_{m=1}^{N_b}  \bC_m^{-1}\bF_{m,I}\right) \bF_{n,I}^T \right] \bC_n^{-1} \right\}^{-1}.
\end{equation}
The signal channel mean is
\begin{equation}
    \bgamma^{(21)} = \bDelta^{(21)} \left\{\sum_{n=1}^{N_b}\sum_{S=\{I,Q,U,V\}}\left[\delta_{SI}\bI - \left( \sum_{m=1}^{N_b} \bC_m^{-1}\bF_{m,I} \right) \bF_{n,S}^T \right] \bC_n^{-1}\by_{n,S} \right\}.
\end{equation}
For computation purposes, we define the total signal noise covariance, $\bC_T$, the weighted total power basis, $\bH$, the weighted total power data, $\bw$, and the overlap vector of the data, $\bd$, through
\begin{equation}
  \bC_T^{-1} = \sum_{k=1}^{N_b}\bC_k^{-1} \text{, } \ \ 
  \bH = \sum_{k=1}^{N_b} \bC_k^{-1}\bF_{k,I} \text{, } \ \ 
  \bw = \sum_{k=1}^{N_b} \bC_k^{-1}\by_{k,I} \text{,  and} \ \ 
  \bd = \sum_{k=1}^{N_b}\sum_{S\in\{I,Q,U,V\}} \bF_{k,S}^T\bC_k^{-1}\by_{k,S}.
\end{equation}
With these definitions, $\bDelta^{(21)}$ and $\bgamma^{(21)}$ are given by
\begin{equation}
  \bDelta^{(21)} = \left(\bC_T^{-1}-\bH\bH^T\right)^{-1} \ \ \text{ and } \ \ \bgamma^{(21)} = \bDelta^{(21)}(\bw-\bH\bd).
\end{equation}
These equations also work when calculating the channel mean and covariance with total power spectra only as long as the sum over $S$ in the definition of $\bd$ includes only $I$.

\section{Signal bias statistic under minimum assumption analysis} \label{app:minimum-assumption-analysis-signal-bias-statistic}

In this appendix, we examine the effect of biases in the foreground model, i.e. the effect of non-zero $\bdelta$ on the uncertainties of minimum assumption analyses. To do this, we write the data as $\by=\bF\bx+\bdelta+\bPsi\bz+\bn$, where $\bdelta$ is the foreground bias (i.e. unmodeled foreground) that satisfies $\bF^T\bC^{-1}\bdelta=\bzero$, $\bF\bx$ is the modeled foreground, $\bz$ is the true signal, and $\bn$ is the Gaussian noise realization with covariance $\bC$. With these definitions, the signal channel mean $\bgamma^{(21)}$ is given by
\begin{equation}
    \bgamma^{(21)} = \left[\bPsi^T\bC^{-1}(\bI-\bPhi)\bPsi\right]^{-1} \bPsi^T\bC^{-1}(\bI-\bPhi)(\bF\bx+\bdelta+\bPsi\bz+\bn).
\end{equation}
Since $\bPhi\bF\bx=\bF\bx$ and $\bPhi\bdelta=\bzero$, this means
\begin{subequations}
\begin{align}
    \bgamma^{(21)} &= \left[\bPsi^T\bC^{-1}(\bI-\bPhi)\bPsi\right]^{-1} \left[\bPsi^T\bC^{-1}\bdelta + \bPsi^T\bC^{-1}(\bI-\bPhi)\bPsi\bz+\bPsi^T\bC^{-1}(\bI-\bPhi)\bn\right], \\
    &= \left[\bPsi^T\bC^{-1}(\bI-\bPhi)\bPsi\right]^{-1}\bPsi^T\bC^{-1}\bdelta + \bz + \left[\bPsi^T\bC^{-1}(\bI-\bPhi)\bPsi\right]^{-1}\bPsi^T\bC^{-1}(\bI-\bPhi)\bn, \\
    &= \bDelta^{(21)}\bPsi^T\bC^{-1}\bdelta + \bz + \bDelta^{(21)}\bPsi^T\bC^{-1}(\bI-\bPhi)\bn.
\end{align}
\end{subequations}
This means that the signal channel bias is
\begin{equation}
    \bgamma^{(21)} - \bz = \bDelta^{(21)}\bPsi^T\bC^{-1}\left[\bdelta+(\bI-\bPhi)\bn\right].
\end{equation}
The so-called signal bias statistic, $\varepsilon^2$, defined through $\varepsilon^2 = \frac{1}{N_\nu}(\bgamma^{(21)}-\bz)^T\left[\bDelta^{(21)}\right]^{-1}(\bgamma^{(21)}-\bz)$, which yields the squared number of sigma at which the signal is measured in an averaged sense across the band, satisfies
\begin{equation}
  \varepsilon^2 =\frac{1}{N_\nu} \left[\bdelta+(\bI-\bPhi)\bn\right]^T\bC^{-1}\bPsi\bDelta^{(21)}\bPsi^T\bC^{-1}\left[\bdelta+(\bI-\bPhi)\bn\right],
\end{equation}
where $N_\nu$ is the number of frequency channels.~Assuming that $\bdelta$ follows a normal distribution (with a singular covariance matrix), the expectation value and variance of $\varepsilon^2$ are
\begin{subequations}
\begin{align}
    \E[\varepsilon^2] &= 1 + \bmu^T\bA\bmu + \Tr(\bA\bSigma)\,, \\
    \Var[\varepsilon^2] &= \frac{2}{N_\nu} + 4\bmu^T\bA\bSigma\bA\bmu + 2\Tr(\bA\bSigma\bA\bSigma) + \frac{4}{N_\nu}\bmu^T\bA\bmu + \frac{4}{N_\nu}\Tr(\bA\bSigma)\,,
\end{align}
\end{subequations}
where $\bA=\frac{1}{N_\nu}(\bI-\bPhi)^T\bC^{-1}\bPsi\bDelta^{(21)}\bPsi^T\bC^{-1}(\bI-\bPhi)$, $\bmu=\E[\bdelta]$, and $\bSigma=\Cov[\bdelta]$.~Assuming that $\varepsilon^2$ approximately follows a normal distribution, this means that, at a confidence level of $p$,
\begin{equation}
  \varepsilon < \sqrt{1 + \bmu^T\bA\bmu + \Tr(\bA\bSigma) + 2\sqrt{\frac{1}{N_\nu} + 2\bmu^T\bA\bSigma\bA\bmu + \Tr(\bA\bSigma\bA\bSigma) + \frac{2}{N_\nu}\bmu^T\bA\bmu + \frac{2}{N_\nu}\Tr(\bA\bSigma)}\ \text{erf}^{-1}(2p-1)}.
\end{equation}
This equation connects the RMS number of sigma, $\varepsilon$, to confidence levels, $p$, in the general case where the foreground model is imperfect.

\section{Taking derivative of finite-dimensional signal with matrix} \label{app:minimum-assumption-analysis-derivative-for-motion-induced-dipole}

In Section~\ref{sec:motion-induced-dipole}, the frequency derivative of a signal $\bs=\begin{bmatrix} s(\nu_1) & s(\nu_2) & \cdots & s(\nu_{N_{\nu}}) \end{bmatrix}^T$ is written as $\bD\bs$. There are some subtleties to this representation. First, the discrete nature of $\bs$ means that any derivative taken will be a finite difference approximation. Second, note that the derivative of a vector with $N_{\nu}$ elements is only strictly defined on the $N_{\nu}-1$ midpoints. The derivative matrix defined in this way is $(N_\nu-1)\times N_\nu$ and in the case of equally spaced frequency channels with resolution $\Delta\nu$ is given by
\begin{equation}
  \bD = \frac{1}{\Delta\nu}\begin{bmatrix} -1 & 1 & 0 & \cdots & 0 & 0 & 0 \\ 0 & -1 & 1 & \cdots & 0 & 0 & 0 \\ 0 & 0 & -1 & \cdots & 0 & 0 & 0 \\ \vdots & \vdots & \vdots & \ddots & \vdots & \vdots & \vdots \\ 0 & 0 & 0 & \cdots & -1 & 1 & 0 \\ 0 & 0 & 0 & \cdots & 0 & -1 & 1 \end{bmatrix}. \label{eq:naive-derivative-matrix}
\end{equation}
To define $\bD\bs$ in the same space as $\bs$, the derivative must be interpolated in some way (i.e. $\bD$ must be made square in some way). A natural interpolation scheme is to take the derivative at the first (last) endpoint to be the derivative at the midpoint of the first (last) pair of elements and to take the derivative at each interior point to be the average of the derivatives at the two adjacent midpoints. Defined in this way, the derivative matrix from Equation~\ref{eq:naive-derivative-matrix} is modified to
\begin{subequations}
\begin{align}
  \bD &=\left(\frac{1}{2}\begin{bmatrix} 2 & 0 & 0 & \cdots & 0 & 0 \\ 1 & 1 & 0 & \cdots & 0 & 0 \\ 0 & 1 & 1 & \cdots & 0 & 0\\ \vdots & \vdots & \vdots & \ddots & \vdots & \vdots \\ 0 & 0 & 0 & \cdots & 1 & 0 \\ 0 & 0 & 0 & \cdots & 1 & 1 \\ 0 & 0 & 0 & \cdots & 0 & 2 \end{bmatrix}\right)\left(\frac{1}{\Delta\nu}\begin{bmatrix} -1 & 1 & 0 & \cdots & 0 & 0 & 0 \\ 0 & -1 & 1 & \cdots & 0 & 0 & 0 \\ 0 & 0 & -1 & \cdots & 0 & 0 & 0 \\ \vdots & \vdots & \vdots & \ddots & \vdots & \vdots & \vdots \\ 0 & 0 & 0 & \cdots & -1 & 1 & 0 \\ 0 & 0 & 0 & \cdots & 0 & -1 & 1 \end{bmatrix}\right), \\
  &= \frac{1}{2\Delta\nu}\begin{bmatrix} -2 & 2 & 0 & \cdots & 0 & 0 & 0 \\ -1 & 0 & 1 & \cdots & 0 & 0 & 0 \\ 0 & -1 & 0 & \cdots & 0 & 0 & 0 \\ \vdots & \vdots & \vdots & \ddots & \vdots & \vdots & \vdots \\ 0 & 0 & 0 & \cdots & 0 & 1 & 0 \\ 0 & 0 & 0 & \cdots & -1 & 0 & 1 \\ 0 & 0 & 0 & \cdots & 0 & -2 & 2 \end{bmatrix}.
\end{align}
\end{subequations}
This is the derivative matrix referred to in Section~\ref{sec:motion-induced-dipole}.

\bibliographystyle{aasjournal}
\bibliography{references}

\end{document}